\documentclass[11pt,a4paper,headinclude,footinclude,chapterprefix=on]{scrreprt}

\usepackage{url}
\usepackage{hyperref}
\usepackage{wrapfig}
\usepackage{units}

\usepackage[printonlyused]{acronym}

\usepackage{calc}
\usepackage{amsmath}
\usepackage{amssymb}
\usepackage{amsfonts}

\usepackage{algorithm}
\usepackage{algpseudocode}
\usepackage{color}
\usepackage[table]{xcolor} 
\usepackage[caption=false]{subfig}
\usepackage{stfloats}
\usepackage{nicefrac}

\usepackage{multirow}
\usepackage[procnames]{listings}

\newcolumntype{C}[1]{>{\centering\arraybackslash}p{#1}}

\usepackage{soul}
\usepackage{booktabs}

\definecolor{BgGray}{gray}{0.7}%
\definecolor{BgGray2}{gray}{0.96}%
\definecolor{RowColorOdd}{named}{BgGray2}%
\definecolor{RowColorEven}{named}{white}%
\definecolor{comments}{gray}{.5}
\definecolor{Gray}{gray}{0.85}
\definecolor{red}{RGB}{160,0,0}
\definecolor{green}{RGB}{0,150,0}
\definecolor{deepblue}{rgb}{0,0,0.5}
\definecolor{deepred}{rgb}{0.6,0,0}
\definecolor{deepgreen}{rgb}{0,0.5,0}

\usepackage{pifont}
\usepackage[multiuser,nomargin,marginclue,inline]{fixme}
\fxusetheme{color}
\fxsetup{targetlayout=colorcb}

\usepackage[utf8]{inputenc}
\usepackage[T1]{fontenc}
\usepackage{tabularx}
\usepackage{multicol}
\usepackage{graphicx}

\usepackage{tikz}
\usepackage{lipsum}

\usepackage[headsepline,plainheadsepline,footsepline,plainfootsepline]{scrpage2}

\usepackage{mathptmx}
\usepackage[scaled=.92]{helvet}
\usepackage{courier}

\usepackage[multiuser,nomargin,marginclue,inline]{fixme}
\fxusetheme{color}
\fxsetup{targetlayout=colorcb}
\FXRegisterAuthor{all}{anall}{ALL}
\FXRegisterAuthor{md}{anmd}{MD}
\FXRegisterAuthor{tolja}{antolja}{Tolja}
\FXRegisterAuthor{mc}{anmc}{Mikolaj}
\FXRegisterAuthor{cp}{ancp}{CP}
\FXRegisterAuthor{awo}{anawo}{AWO}
\FXRegisterAuthor{pg}{anpg}{Piotr}

\fxsetup{status=draft}

\usepackage{geometry}
\clearscrheadfoot

\ihead[\Large {\scshape TU Berlin }]{\Large {\scshape TU Berlin }}

\ifoot[{\tiny \begin{minipage}{4.0cm}Copyright at Technische Universität Berlin.\newline All Rights reserved.\end{minipage}}]{{\tiny \begin{minipage}{4.0cm}Copyright at Technische Universität Berlin.\newline All Rights reserved.\end{minipage}}}
\ofoot[Page \pagemark]{Page \pagemark}

\addtokomafont{pagefoot}{\normalfont}

\graphicspath{{figures/}}
\newcommand{\trnumber}{TKN 16-0003} 
\newcommand{\trdate}{November 2016}
\newcommand{\trauthor}{Piotr Gawłowicz, Anatolij Zubow, Mikolaj Chwalisz and Adam Wolisz}
\newcommand{\tremail}{\{gawlowicz, zubow, chwalisz, wolisz\}@tkn.tu-berlin.de}
\newcommand{\trtitle}{UniFlex: Accelerating Networking Research and Experimentation through Software-Defined Wireless Networking}

\begin{document}


{
\sffamily

\thispagestyle{empty}

\setlength{\tabcolsep}{0pt} 
\noindent 
\begin{tabularx}{\columnwidth}{cXc}
  \includegraphics[height=1cm]{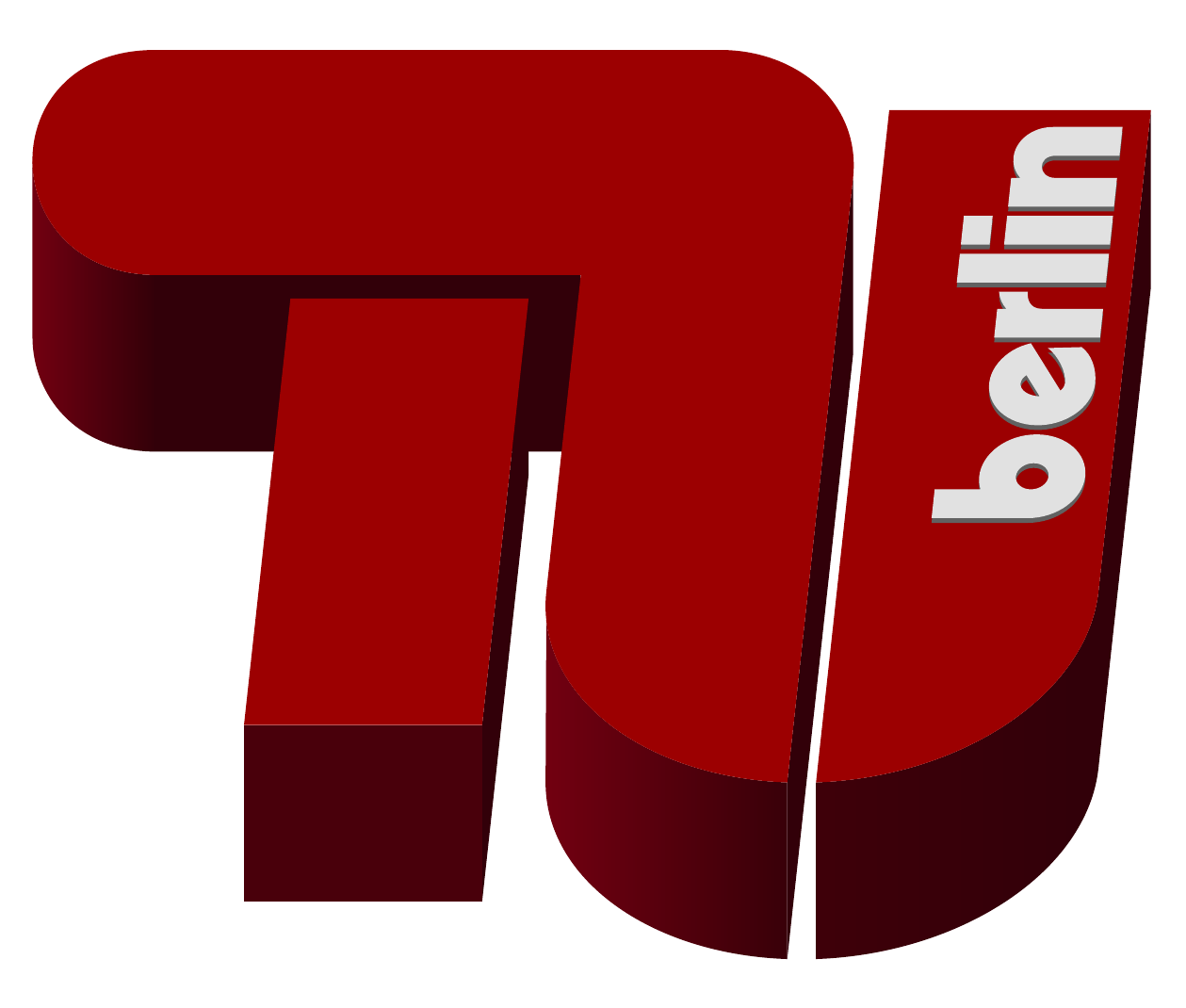}
  & &
  \includegraphics[height=1cm]{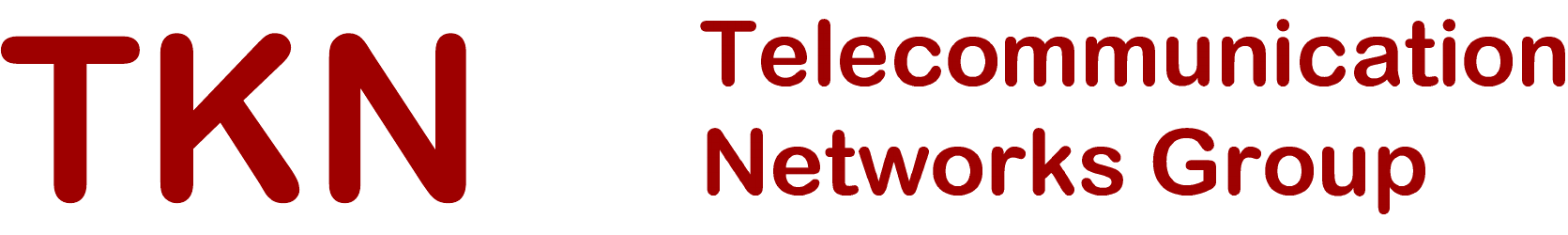}
  \\
\end{tabularx}
\setlength{\tabcolsep}{6pt} 

\vspace{1.0cm}

\begin{center}
{\huge
\noindent
Technische Universität Berlin

\vspace{0.5cm}

\noindent
Telecommunication Networks Group

\begin{center}
\rule{15.5cm}{0.4pt}
\end{center}
}
\end{center}

\begin{minipage}[][11.0cm][c]{14.5cm}
{\Huge

\begin{center}
\trtitle
\end{center}

\begin{center}
{\LARGE \trauthor} \\
{\Large \tremail}
\end{center}

\begin{center}
Berlin, \trdate
\end{center}

\vspace{0.5cm}

}

\begin{center}
\setlength{\fboxrule}{2pt}\setlength{\fboxsep}{2mm}
\fbox{TKN Technical Report \trnumber}
\end{center}

\end{minipage}

\setlength{\fboxrule}{0.4pt}
\setlength{\fboxsep}{0.4pt}

\begin{center}

  \rule{15.5cm}{0.4pt}

  \vspace{0.5cm}

  {\huge {TKN Technical Reports Series}}

  \vspace{0.5cm}

  {\huge Editor: Prof. Dr.-Ing. Adam Wolisz}

  \vspace{0.5cm}

 \end{center}

}


\begin{abstract}
\subsection*{\abstractname}
lassical control and management plane for computer networks is addressing individual parameters of protocol layers within an individual wireless network device. We argue that this is not sufficient in phase of increasing deployment of highly re-configurable systems, as well as heterogeneous wireless systems co-existing in the same radio spectrum which demand harmonized, frequently even coordinated adaptation of multiple parameters in different protocol layers (cross-layer) in multiple network devices (cross-node).

We propose UniFlex, a framework enabling unified and flexible radio and network control. It provides an API enabling coordinated cross-layer control and management operation over multiple network nodes. The controller logic may be implemented either in a centralized or distributed manner. This allows to place time-sensitive control functions close to the controlled device (i.e., local control application), off-load more resource hungry network application to compute servers and make them work together to control entire network.

The UniFlex framework was prototypically implemented and provided to the research community as open-source. We evaluated the framework in a number of use-cases, what proved its usability.
\end{abstract}

\tableofcontents



\chapter{Introduction}

The control plane and the management plane have played a very important role in the classical telecommunication systems, but have been given much less attention in computer networks. As the matter of fact the only widely accepted approach is the usage of SNMP or NETCONF as basis for creating management applications. This is increasingly recognized as not sufficient - especially in case of wireless networks where many parameters have to be frequently tuned in response to changing wireless propagation, interference and traffic conditions. There were already couple of attempts for wireless control protocols including LWAPP and CAPWAP, but those were designed with focus on configuration and device management and are not suitable for time-sensitive control of devices.

Furthermore, classical control/management actions have been addressing individual parameters of protocol layers within an individual network device. This is not sufficient in phase of increasing deployment of highly re-configurable systems, as well as heterogeneous wireless systems co-existing in the same radio spectrum which demand harmonized, frequently even coordinated (simultaneous) change of multiple parameters in different parts of hardware and software in multiple network devices. 

Typical example of emerging real scenarios are LTE-U and Wi-Fi in 5\,GHz and Wi-Fi, Bluetooth and ZigBee in 2.4\,GHz ISM band. On the other hand even homogeneous deployments are suffering from intra-technology interference~\cite{mvulla2015analysis}. Here in recent years we have seen a boom of cross-layer design proposals for wireless networks~\cite{pejovic2014whiterate,kumar2013bringing} where additional information from some layers are obtained and used to optimize operation of another layer.

So far control applications had to solve the challenges of harmonized/simultaneous actions on case-by-case basis, which significantly complicated development of such applications - and lead to lack of any compatibility across the individual solutions. 
We argue that the efficiency of wireless networks can be significantly improved by enabling the management and control of the different co-located wireless technologies and their network protocols stacks (cross-layer) in a coordinated way using either centralized or distributed controllers~\cite{ryuo}. 

\medskip

\textbf{Contribution:} In this paper we propose \textbf{UniFlex}, a framework for \textbf{Uni}fied and \textbf{Flex}ible radio and network control. The suggested API supports typical functions needed for coordinated cross-layer, cross-technology and cross-node control and similarly as in the SDN paradigm we allow for centralized control, but support equally well also hierarchical control structure and logically centralized but physically distributed control.
Network control applications can be either co-located with controlled device (both running on the same network node, e.g. for latency reasons) or separated from each other (running on two nodes, e.g. control application runs on server due to requiring high computing power).
We believe that UniFlex will be an enabler for rapid prototyping of control applications for wireless network devices, management  and control of operation of - possibly heterogeneous - nodes in wireless networks.
The UniFlex prototype is provided as open-source to the community: https://github.com/tkn-tub/UniFlex.

\medskip

The rest of this report is organized as follows. In Sec. II the system model is introduced, while in Sec. III requirements for the wireless network control plane are defined. Sec. IV describes the architecture of envisioned UniFlex framework. In Sec. V and VI details of our prototypical implementation and integration with external software are given. Examples of wireless network control applications implemented using UniFlex are presented in Sec. VII. Performance of UniFlex prototype is evaluated in Sec. VIII. In Sec. IX related research is discussed and finally, Sec. X summarizes our main findings and concludes the paper.

\chapter{System Model}

In this section, we define our system model and provide definitions of all terms that we use consistently in this paper. 

The \textbf{network} is a collection of \textbf{nodes} under common management and control domain (authority). A \textbf{node} is collection of equipment sharing a common platform and being run under a single instance of operating system. The types of nodes span from small constrained devices to powerful compute servers.
A node is equipped with zero or more \textbf{devices}.
A \textbf{device} is a piece of hardware fulfilling a dedicated functionality. Additionally, it may expose set of operations in Native Device Programming Interface (\textbf{NDPI}) to control it's behaviour and parameters. There are different classes of sets of operations for different devices. For example, a \textbf{wireless network device} provides packet forwarding functions with usage of wireless transmission technology (e.g. 802.11, LTE) and exposes NDPI to control its parameters including transmission power, central frequency, bandwidth, etc.


Operation of a network is harmonized by single or multiple \textbf{controllers}. The controller logic may be implemented either as standalone or multiple cooperating \textbf{control applications} (aka network applications) that run in node(s). In particular, a control application may be located in the same node as network device that it is controlling. 

We assume the existence of a common \textbf{Control Channel} enabling control application(s) to: i) access NDPI of all devices in network, ii) use it to control their behaviour and iii) exchange control messages between each other for cooperation purposes. This control channel may be realized over the  wireless network itself (in-band) and/or additional wired backhaul infrastructure (out-band).

\begin{figure}[!ht]
   \begin{center}
       \includegraphics[width=0.6\linewidth]{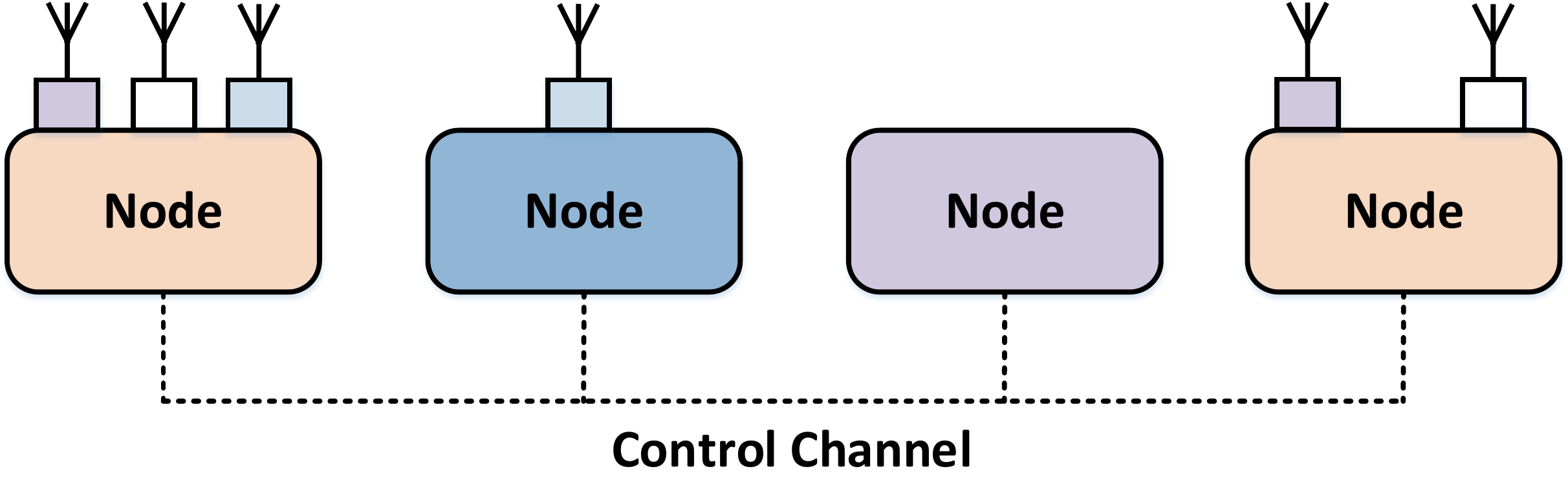}
   \end{center}
    \vspace{-10pt}
   \caption{System model overview.}
   \label{fig:system_model}
\end{figure}


\chapter{Requirements and Design Principles}

\begin{figure}[!ht]
   \begin{center}
       \includegraphics[width=0.7\linewidth]{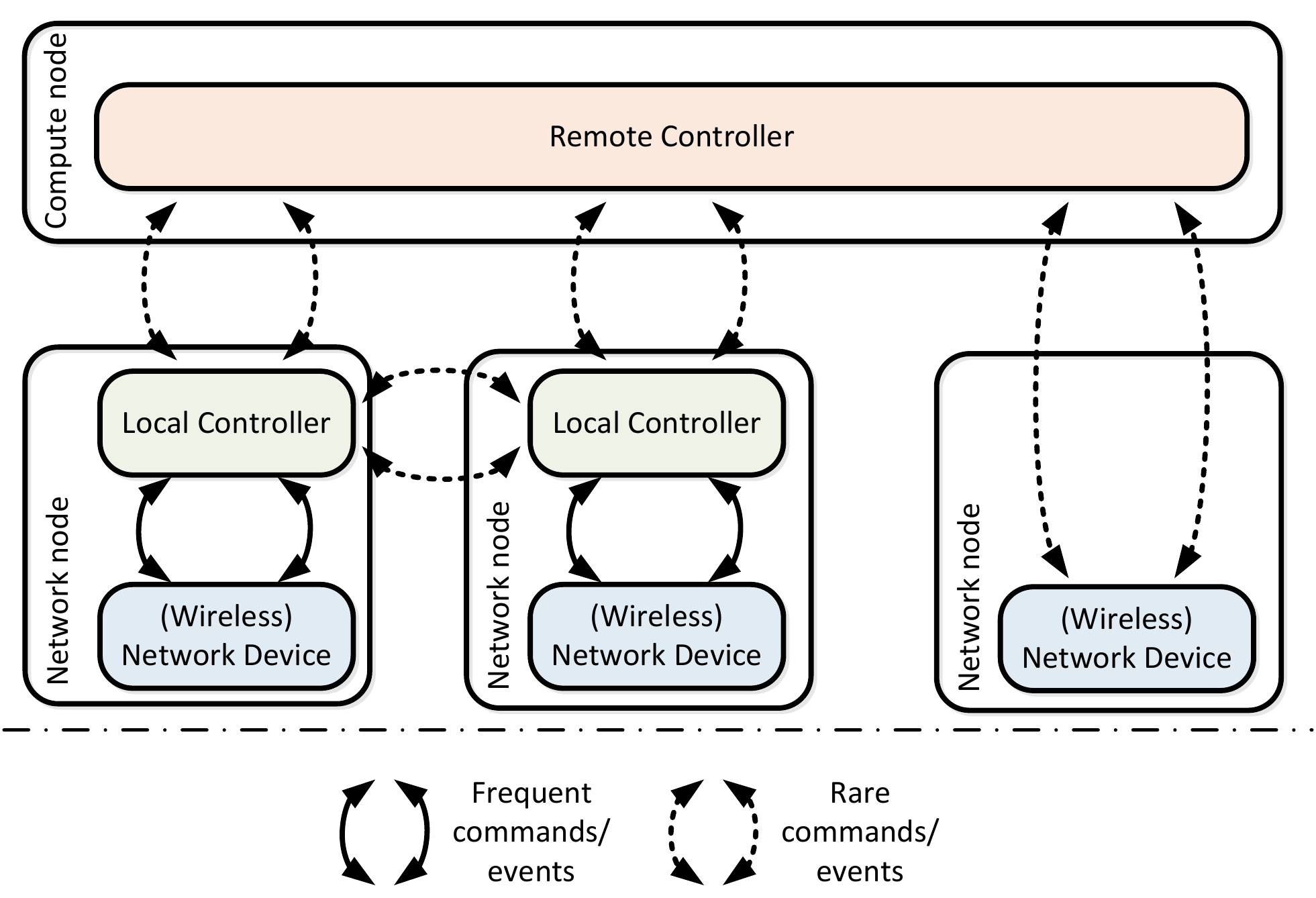}
   \end{center}
    \vspace{-10pt}
   \caption{UniFlex's levels of control. Local controllers handle frequent commands and events, while remote controllers handle rare events. The exchange of events between controllers is also rare.}
   \label{fig:uniflex_level_of_ctrls}
\end{figure}

The main goal of of our work is to facilitate and shorten time required for prototyping of novel control solutions in heterogeneous wireless networks. We argue that novel wireless applications may be realized when following functionality is provided:
\begin{itemize}
    \item Coordinated \textbf{collection} of information from and \textbf{execution} of commands on different protocol layers (\textbf{cross-layer}), heterogeneous devices (\textbf{cross-technology}) and multiple nodes (\textbf{cross-node}) within network,
    \item Existence of a global and consistent view of the entire network, i.e. knowledge about the state of all devices and their relationship,
    \item Possibility to implement logically centralized and physically distributed control applications, i.e. placing time-sensitive task close to device and off-loading greedy tasks to powerful servers,
    \item Support for multiple levels of control for scalability reasons, i.e. local control applications handle frequent commands and events, while remote controllers handle rare events (Fig.~\ref{fig:uniflex_level_of_ctrls}),
    \item Support for detecting network changes in proactive and reactive control schemes in control applications,
    \item A high-level API for control of operation of individual wireless devices and groups of devices,
    \item Location transparency i.e. the same API syntax for execution of commands on local and remote devices,
    \item Possibility to execute commands on group of nodes/devices.
\end{itemize}

\medskip

Control and optimization of operation of wireless network usually involves tuning parameters of network devices being in proximity of each other, i.e. in wireless communication/interference/sensing area. Examples are the radio channel and transmit power assignment to co-located Access Points in Wi-Fi networks. Hence, the control plane requires mechanism to discover the wireless devices in the network and their (wireless) relationship. Moreover, this information has to be monitored and updated at run-time. 

Having a global view on the entire wireless network enables control programs to efficiently manage and control of wireless devices. Changes in the network state can be detected in two ways, namely proactive and reactive. In proactive approach, the 
network controller is polling the network entities, while in reactive approach execution of control program functions is triggered by events generated by the nodes in the network. It should be up to experimenter to define her control strategy. 

For coordinated control among multiple devices of different nodes, the framework API has to support time synchronized execution of functions across multiple network devices. Examples are the coordinated channel switch of multiple devices due to appearance of an interference source.

While it is natural that the device programming interface is different for each wireless technology, in most cases it also varies across different implementation of the same technology, i.e. wireless devices of different vendors. A unification of the different Native Device Programming Interface (NDPI) would be desirable as it would allow controlling the devices of a HetNet in a unified way. Similar concept is nowadays developed for wired switches, e.g. Switch Abstraction Interface~\cite{SAI}.

In the general SDN concept the control plane is logically centralized enabling control programs to have a global view of the entire network. This approach simplifies the development of control application a lot. However, from the practical point of view a centralized controller would introduce a significant delay in the control plane, what can prevent time sensitive control logic to be implemented. Moreover, transporting all monitoring data from devices to a central controller may create too high load on the control plane. Sometimes pre-processing data locally at the node is feasible. In this way, controller may be partitioned into smaller applications where parts of them would run on the network nodes and others on the central compute node. Another advantage of such a split is the possibility of re-usage of control applications. For example, an averaging filter may be implemented once as control application and used as part of many other controllers. Fig.~\ref{fig:uniflex_level_of_ctrls} shows the UniFlex's levels of control. Local controllers handle frequent commands and events, while remote controllers handle rare events. The exchange of events between controllers is also rare.

In summary, the envisioned control framework may be considered as a wireless network operating system, that runs multiple control applications (network apps) communicating with each other and providing them with an interface for controlling all wireless network devices in coordinated way.


\chapter{Architecture Overview}
\label{sec:architecture}

\begin{figure}[!ht]
   \begin{center}
       \includegraphics[width=0.65\linewidth]{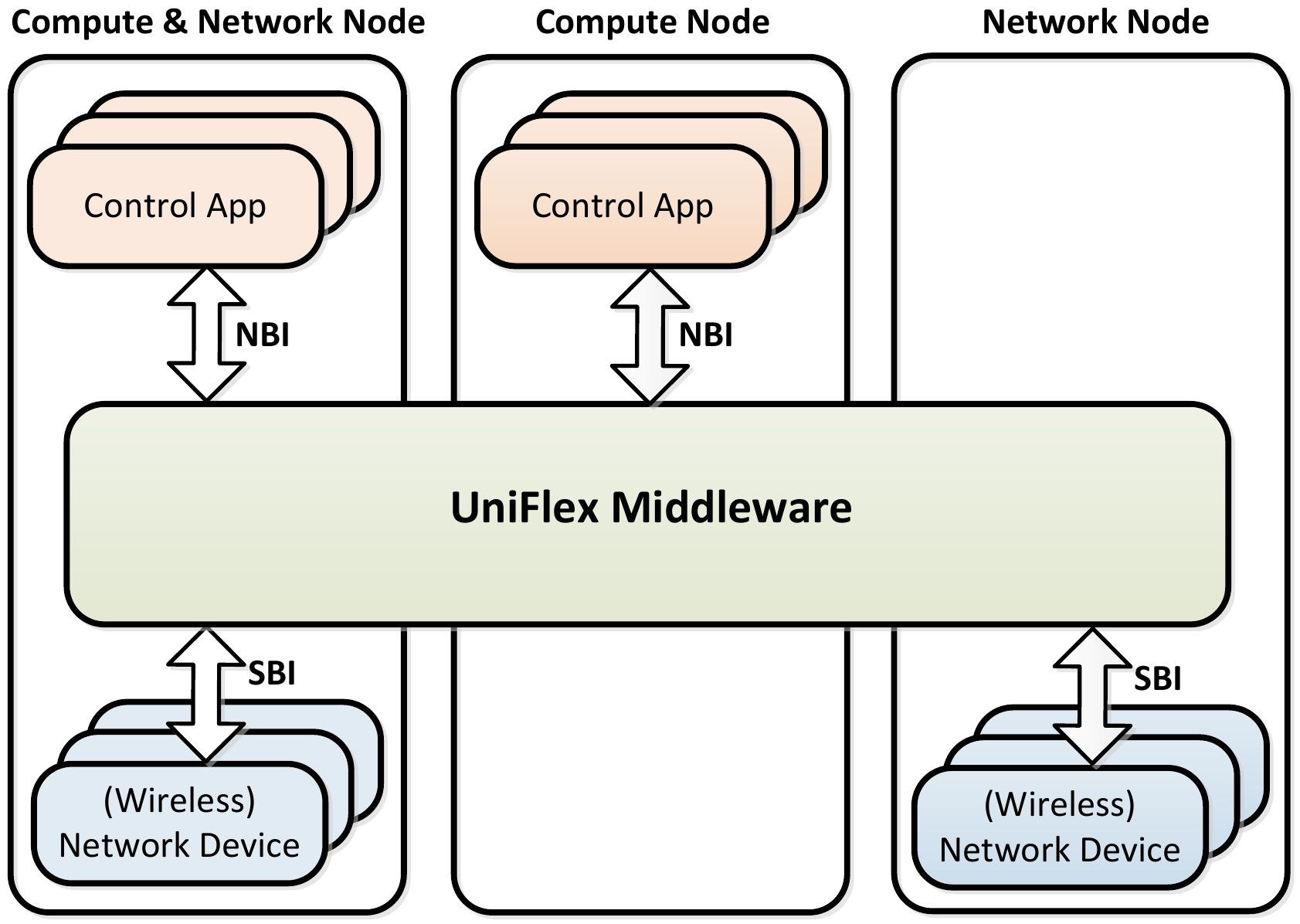}
   \end{center}
    \vspace{-10pt}
   \caption{UniFlex's high-level architecture: Northbound Interface (NBI) and Southbound Interface (SBI).}
   \label{fig:framework_architecture}
\end{figure}

Fig.~\ref{fig:framework_architecture} shows an overview of the architecture of the UniFlex framework. The framework is a distributed middleware platform running across multiple nodes that interconnects all control applications and wireless network devices. The network applications running on top of middleware perform control tasks over wireless devices by utilizing the provided API in Northbound Interface. The Southbound Interface is responsible for translating calls coming from control applications to NDPI.
In the following subsections, we provide detailed description of the UniFlex framework. 

%
%
\section{Control Applications}

A control application is an entity that implements particular network controller logic. In general, it collects information and measurements from network, make control decisions according to set policy and perform network reconfiguration. 

Each control application is provided with a global view on all nodes in the network. Basically, control application is operating on proxy object of remote nodes. Each node proxy contains proxy objects referring to application and device modules present in a given remote node. 
By default, each control application is able to communicate with all applications and control all device modules in entire wireless network, but some access policies may be applied.

Control applications may send as well as receive events generated by framework, devices and other applications. In order to be notified about specific event, an application has to subscribe to this event. 
Events originated from devices are usually used in reactive control approach, e.g. every time a network device is not able to receive a frame correctly, it sends a \emph{FrameLostEvent}. 
Applications may use events to communicate with each other. 

Events may be sent in three modes: i) unicast, ii) node-broadcast and iii) global-broadcast.
Using unicast mode an application is able to send events to a particular recipient. In node-broadcast mode application sends events to a particular node and all applications running on that given node will receive it. Finally, using global-broadcast applications and devices are able to send events to all applications in the entire network. 
An event always contains an identifier of node and entity (application or device) that generated it. In addition, events of specific type may contain additional data, as measurement samples, etc.

%
%
\section{Northbound Interface}\label{sec:api}

Fig.~\ref{fig:nbi_metamode_short} shows the Northbound Interface description as UML class diagram. Each network control application
operates on proxy objects referring to nodes, applications and device modules. 
Those proxy objects provides an access to the UniFlex Northbound Interface. 
The NBI is the same for devices located in the same host machine as well as for remote devices. It is the framework that is responsible for delivering and executing function calls in proper device.

The Northbound Interface gives an application possibility to send, subscribe for and receive events. An application sends events using \emph{send\_event(event)} function and subscribe for events using \emph{subscribe\_for\_events(eventType, callback)}. Once subscribed the framework will deliver events of proper type to application and fire bounded callback passing event as the argument. 
In order to send/receive events in unicast, node-broadcast and global-broadcast mode, an application has to execute \emph{send\_event()}/\emph{subscribe\_for\_events()} function on application (device) proxy, node proxy and self-object, respectively.

In addition, a control application may execute Remote Procedure Calls (RPC) on device/protocol proxy object to call functions from its native programming interface. By default, all RPC calls are executed on the device/protocol proxy object are blocking execution of control application until function returns.
For convenience, the \texttt{DeviceProxy}/\texttt{ProtocolProxy} class provides three functions, namely: i) \emph{delay(relative\_time)}, ii) \emph{exec\_time(absolute\_time)} and iii) \emph{callback(cb=None)}. Using those functions one may delay execution of call, schedule execution of call in future point in time and execute non-blocking call, respectively. Optionally, it is possible to register callback function for value returned from function call. Calling examples are given in Sec.~\ref{sec:nb_impl}.

In order to control a node, an application has to first obtain a \texttt{NodeProxy} object. This is achieved by subscribing to \texttt{NewNodeEvent} events. On discovery of a new node UniFlex notifies the application about this by sending event containing a node proxy. Thereafter, the application can retrieve the \texttt{DeviceProxy}, \texttt{ProtocolProxy} and/or \texttt{ApplicationProxy} objects from obtained \texttt{NodeProxy} and execute commands on them.
Note that application is notified about all nodes in network including the one that it run on top of -- a proper flag \textit{Local} in \textit{NodeProxy} is used to distinguish between local and remote nodes. 
The framework tracks presence of all nodes in network and notifies applications about node lost events.

%
%
\section{Distributed Control Framework}

The UniFlex is a distributed middleware platform that inter-connects network devices and control applications. 
The framework takes care of node management including node discovery and monitoring connection to between all nodes. Whenever new node is discovered or connection with some node is lost, UniFlex notifies all applications about the changes by sending proper event.

While Control Application running on top of the framework are able to control wireless devices/protocols with simple API, it is the framework, that is responsible for delivering and executing function calls in proper device/protocol.
Moreover, the middle-ware is also responsible for discovery of capabilities of each device/protocol and throwing exceptions if not supported function is to be called. 

Using UniFlex it is possible to develop distributed network control applications, i.e. it is possible to split-up control logic into smaller cooperating control applications running on different nodes.
In this way, UniFlex by design support three types of control programs: i) local -- when controller is running in the same node as controlled device(s), ii) non-local -- when controller is running on different node then controlled device(s) and iii) hybrid or hierarchical -- when controller logic is split between multiple nodes.
A hierarchical control is a trade-off between local and global control. It allows putting time sensitive control functions close to device, and offload complex tasks to remote compute nodes (servers in cloud). 
Note that framework provides location transparency meaning that calling syntax is always the same for local as well as remote device.

%
%
\section{Southbound Interface}\label{sec:sb_interface}

The Southbound Interface works in two directions, i.e. control application may execute functions and change parameters of device (in-direction), but it may also receive data, measurements, samples, etc. from device (out-direction). All communication in in-direction is realized with RPC calls, while communication in out-direction is realized using events.

\begin{figure}[htbp]
   \begin{center}
       \includegraphics[width=0.7\linewidth]{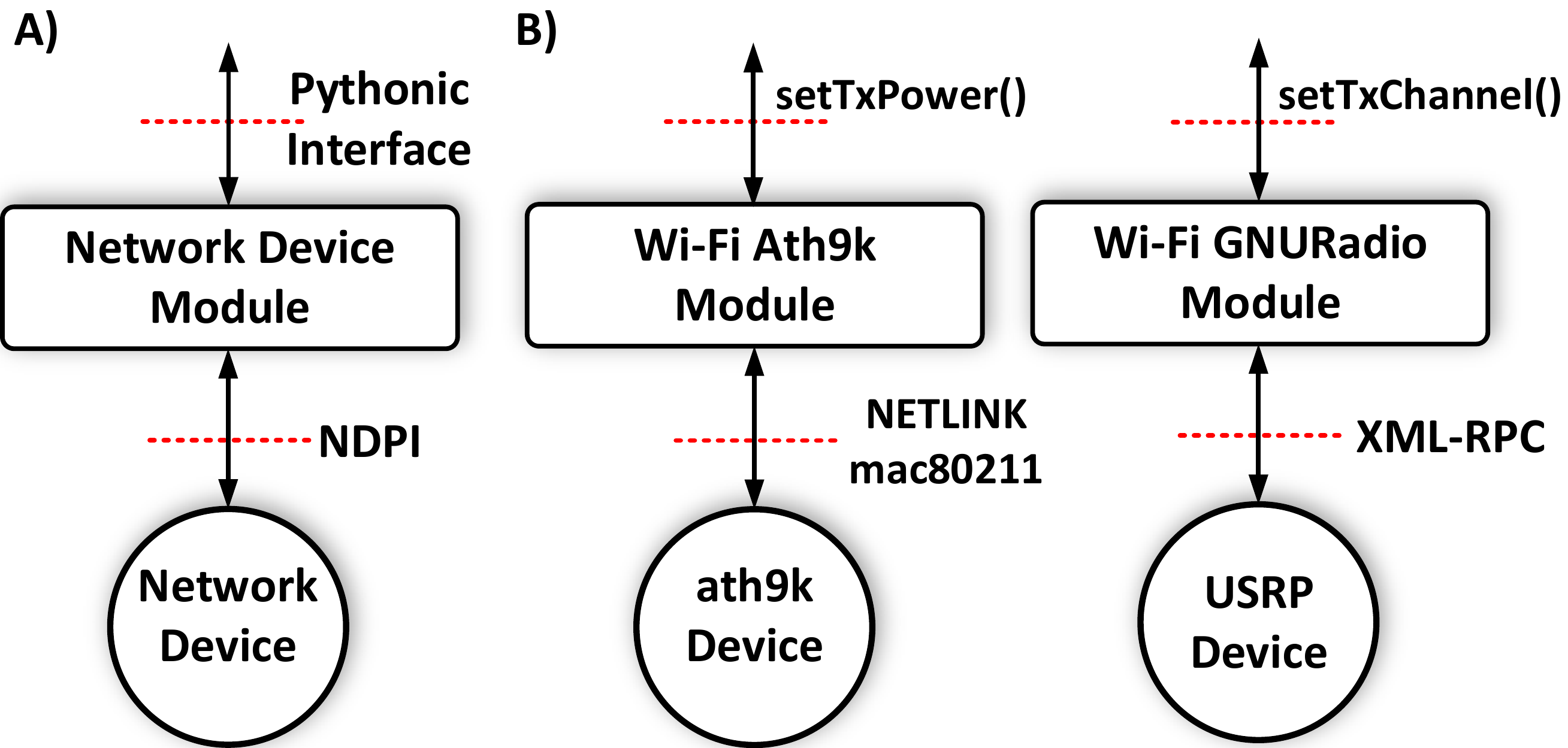}
   \end{center}
    \vspace{-10pt}
   \caption{Device Module is an Python wrapper for Native Device Programming Interface.}
   \label{fig:native_device}
\end{figure}

The Southbound Interface is realized with help of Device and Protocol Modules.
Device Module translates function calls from control applications into Native Device Programming Interface (NDPI) -- Fig.\ref{fig:native_device}A. While Protocol Module translates function calls into Native Protocol Programming Interface (NPPI).
In other words, Device/Protocol Module wraps different API and tools used to program device/protocol, and exposes them to framework in \emph{Pythonic} way (i.e. as Python functions).
In Fig.\ref{fig:native_device}B, we present two Device Modules as an example. As shows the Python functions are delivered to modules and translated to proper NDPI calls, i.e. NETLINK and XML/RPC respectively.

In many cases, while the semantics of features supported by different devices may be the same, their NDPI may differ greatly. We argue that this would be error prone and would prevent portability and re-usibility of control application.
We believe that those issues can be solved by a unified abstraction layer that hides specific NDPIs of different devices behind common one. We found Unified Programming Interface (UPI) defined in WiSHFUL project~\cite{ruckebusch2016unified,fortuna2015wireless} to be an appropriate option.

\section{UniFlex Framework UML Diagram}\label{sec:uniflex_uml}

\begin{figure}[!ht]
   \begin{center}
       \includegraphics[width=1\linewidth]{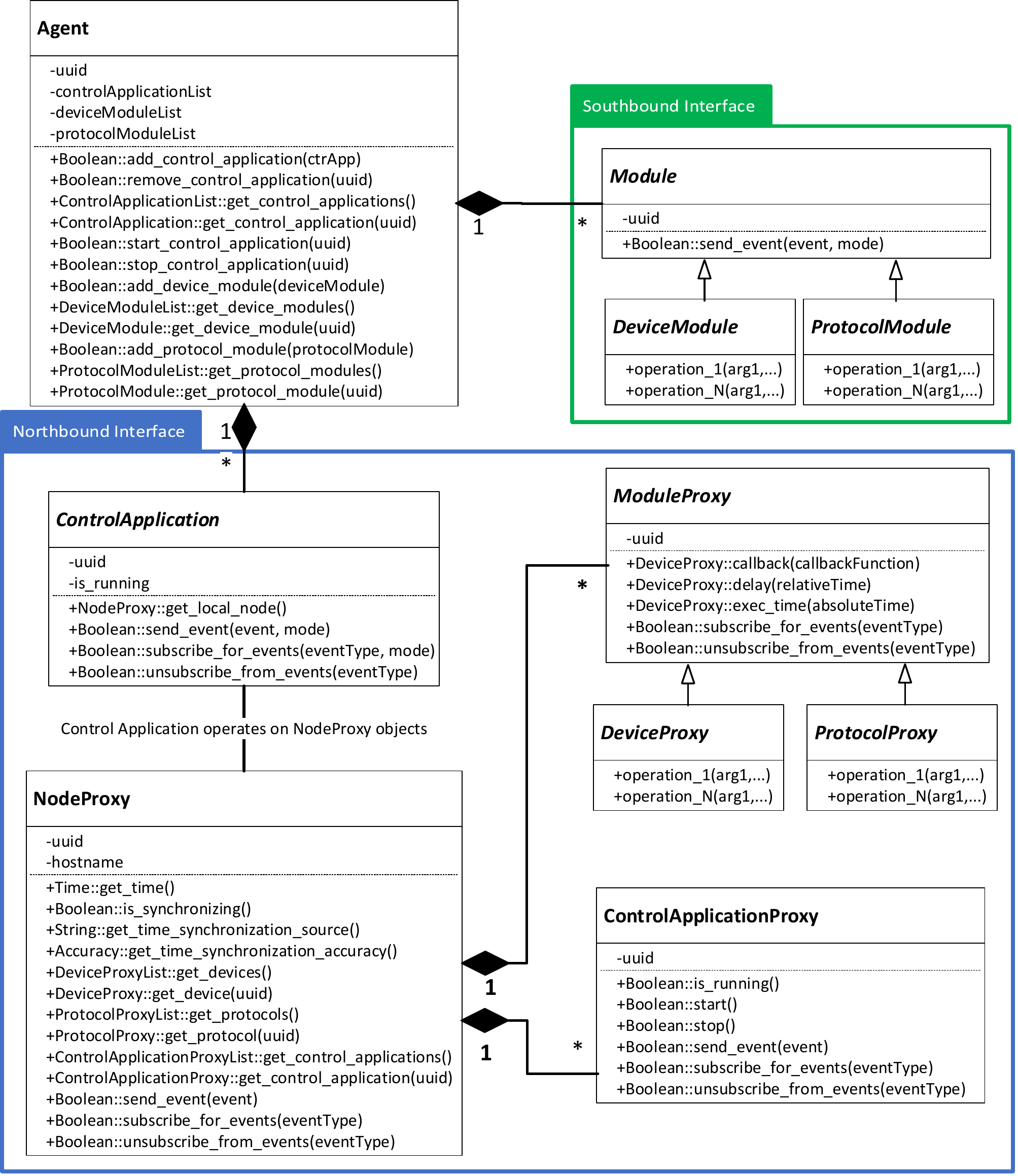}
   \end{center}
    \vspace{-10pt}
   \caption{UniFlex framework (UML class diagram).}
   \label{fig:nbi_metamode_short}
\end{figure}

The UML diagram in Fig.~\ref{fig:nbi_metamode_short} presents UniFlex architecture in more detail. The Distributed Control Framework is consists of set of communicating Agents. The \texttt{Agent} is provides communication middleware for Control Applications, Device and Protocol Modules. Description of functions is given in Appendix A.

\chapter{Implementation Details}

The UniFlex framework was prototypically implemented. Particular attention was paid to enabling code re-usability and support for different programming languages as well as possibility of usage of specialized external software libraries.

Our prototype is implemented in Python language, what makes it possible to run on multiple different host types (Linux, OpenWRT, Mac OS and Windows) and allows for rapid prototyping of control applications. As we used only standard and common Python libraries, we are able to run and test our implementation on multiple platforms, including x86, ARM and MIPS. 

\begin{figure}[!ht]
   \begin{center}
       \includegraphics[width=0.55\linewidth]{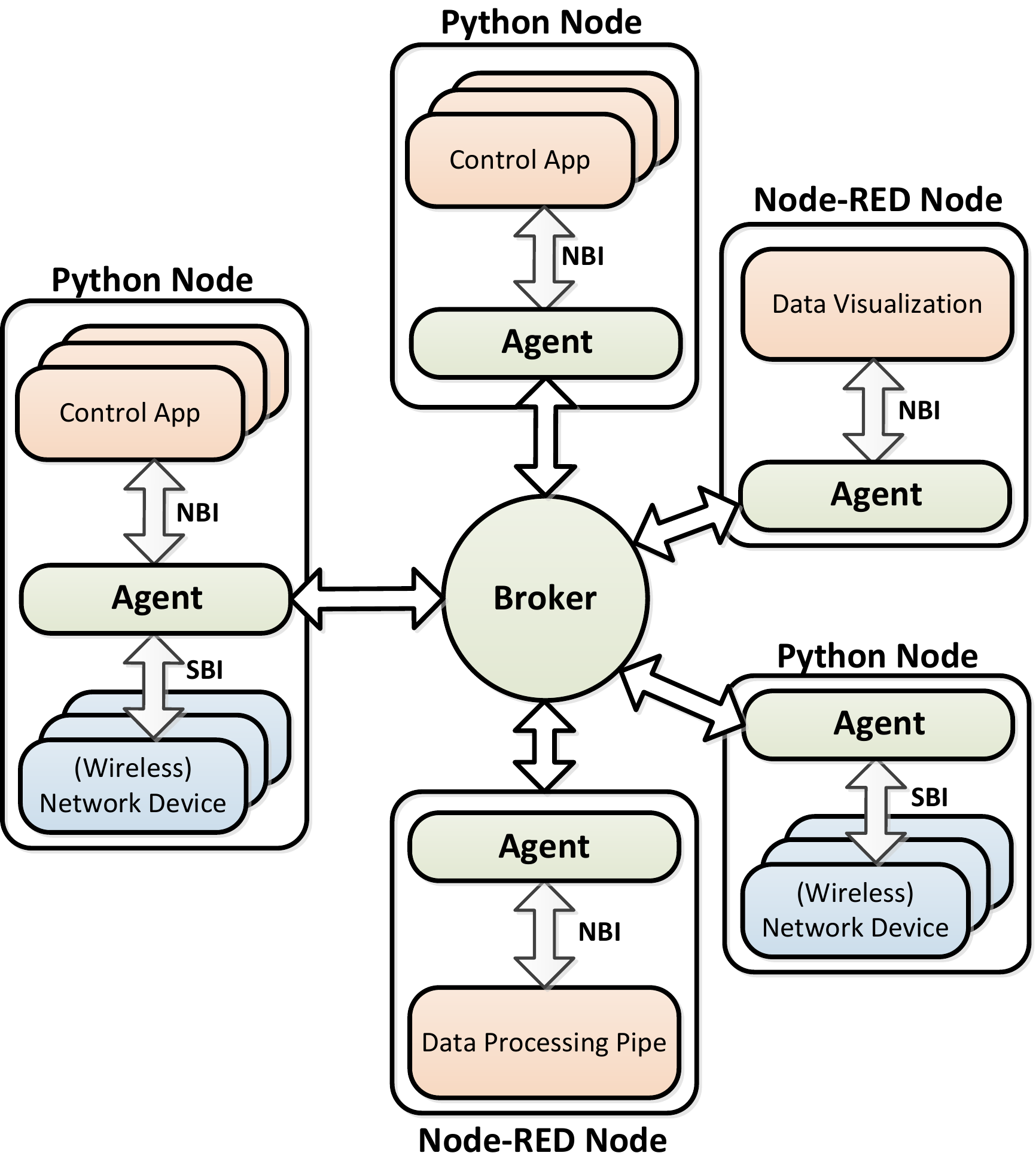}
   \end{center}
    \vspace{-10pt}
   \caption{Overview of UniFlex framework implementation.}
   \label{fig:framework_impl}
\end{figure}

An overview of framework implementation is presented in Fig.\ref{fig:framework_impl}. 
An Agent exposes \emph{Northbound Interface} to Applications and connects to Device Modules using \emph{Southbound Interface}. Agents communicate with each other over Broker.
Each depicted entity will be described in the following subsection in more detail. 

The framework is distributed as a single Python package containing implementation of Agent and Broker, and base classes for Applications and Device Modules. In other words, we provide core functionality of distributed control plane that is able to run on variety of nodes including constrained devices, and leave implementation of control applications and device modules to researchers.

%
%
\section{Northbound Interface}\label{sec:nb_impl}

Sending and subscription of events is implemented around \texttt{PUB} and \texttt{SUB} sockets available in \textit{\O{}MQ} library~\cite{zeromq-2014} communication library. 

The Remote-Procedure-Call mechanism was implemented on top of unicast event mechanism. For this purpose, we introduced two events, namely: \texttt{CommandEvent} and \texttt{ResponseEvent} that are handled internally by Agent. 
The \texttt{CommandEvent} event contains Calling Context, i.e. information \textit{WHAT} (operation), \textit{WHERE} (node and device), \textit{HOW} (blocking, delayed, etc.) and optionally \textit{WHEN} (point in time) is to be executed, while the \texttt{ResponseEvent} contains return value of function call.
Creation of Calling Context and sending of \texttt{CommandEvent} is hidden behind execution of RPC call on device proxy objects.
To facilitate building of Calling Context, we introduce following functions: \textit{delay()}, \textit{exec\_time()} and \textit{callback()} that may be executed on device proxy object. They are optional and may be chained together.
The examples of the supported calling semantics are presented in Listing~\ref{list:rpc_call}.

In order to support delayed and time-scheduled function execution, the Agent class is equipped with scheduler (Python Apscheduler).
Note, when coordinating multiple nodes by means of time scheduled execution, nodes in network must have common notion of global clock (e.g. GPS or time protocols like PTP).

\lstset{language=Python,
        basicstyle=\ttfamily\footnotesize,
        keywordstyle=\color{deepblue},
        commentstyle=\color{comments},
        stringstyle=\color{deepgreen},
				emphstyle=\ttb\color{deepred},    
        showstringspaces=false,
        procnamekeys={def,class}}

\begin{lstlisting}[caption=Calling examples., label=list:rpc_call]
#definition of callback function
def my_get_power_cb(data):
    print(data)

#get device proxy from node proxy
device = node.get_device(0)

#execution of blocking call
result = device.radio.get_tx_power()

#execution of non-blocking call
device.callback(my_get_power_cb).radio.get_tx_power()

#delay execution of non-blocking call
device.delay(3).callback(my_get_power_cb).radio.get_tx_power()

#schedule execution of non-blocking call
t = datetime.now() + timedelta(seconds=3)
device.exec_time(t).callback(my_get_power_cb).radio.get_tx_power() 
\end{lstlisting}

%
%
\section{Agent}

The Agent is an entity that runs Applications and Device Modules, and connects them to distributed control plane. 
It is mainly responsible for transferring events between local and remote (over Broker) applications and device modules. Moreover, it implements RPC mechanism as described in previous subsection. 

Agent provides up-to-date information about all nodes in network to all its application. 
For this purpose we implemented discovery and heartbeat (based on hello messages) mechanisms. Every time an Agent discovers new node sending hello messages over Broker, it sends node unicast \emph{NodeInformationRequest} and based on received \emph{NodeInformationResponse} builds \emph{NodeProxy} object. Then, the Agent notifies all its applications using \texttt{NewNodeEvent}. 
From this point, the Agent monitors presence of node by means of refreshing hello timer. If it expires, Agent sends \emph{NodeLostEvent} to all its application. 

Each application and device module is started within its own thread what increase performance, while preventing race-conditions.


%
%
\section{Broker}
While Agent is responsible for transporting events between Applications and Device Modules in single Node, the Broker takes care for delivering events between Nodes. For optimization, events of specific type are delivered only to those nodes that subscribed for them (i.e. listen on topic that is event's name). All Agents together with Broker constitute distributed UniFlex control framework middle-ware.

The Broker is wrapping and switching events between \texttt{XPUB} and \texttt{XSUB} sockets available in \textit{\O{}MQ} library~\cite{zeromq-2014} library. The ZMQ itself implements mechanisms for topic agreement and message routing. 

By default Pickle library is used for (de)serialization of events. This solution is very convenient, because developer do not have to care about writing parsing and serializing functions, which is wearisome and error prone task. 
On the other hand, it allows only for inter-Python communication and has lower performance comparing to fast serialization libraries. 
For inter-language communication and efficiency, we provide a possibility to define parse and serialization functions. If provided, they are used instead of Pickle.
Those functions may be implemented in any format, but it has to be consistent among all nodes. 
Every time an agent sends/receives an event, it checks whether those functions are provided. If so, it uses them, otherwise Pickle functions are called.

%
%
\section{Control Application}

A Control Application is an entity that implements the entire or part of the network control logic. It can be as simple as a signal filter or as complicated as a mobility management unit. Application has to inherit from \texttt{ControlApplication} base class provided in UniFlex framework package.

While it is possible to subscribe at run-time to events using \textit{subscribe\_for\_event()} as described in Sec.~\ref{sec:api}, we expect that mostly permanent (i.e. lasting for life-cycle of an application) subscriptions will be used. 
For this purpose, we provide \textit{on\_event} decorator that binds event notification with desired function. For example in Listing~\ref{list:rpc_call2}, the application subscribes for \texttt{NewNodeEvent} events and gets notification whenever a new node is discovered. 
Each event contains a proxy object to source node and entity (application or device module), that an application may use to perform RPC calls and to send events.

\lstset{language=Python,
        basicstyle=\ttfamily\footnotesize,
        keywordstyle=\color{deepblue},
        commentstyle=\color{comments},
        stringstyle=\color{deepgreen},
				emphstyle=\ttb\color{deepred},    
        showstringspaces=false,
        procnamekeys={def,class}}

\begin{lstlisting}[caption=Example of blocking RPC call, label=list:rpc_call2]

@on_event(NewNodeEvent)
def add_node(self, event):
    node = event.node
    device = node.get_device(0)
    txPower = device.radio.get_tx_power()

\end{lstlisting}

%
%
\section{Southbound Interface}




%
%

The Southbound Interface is realized with help of Device Modules that translate function calls from control applications into Native Device Programming Interface (NDPI) and Protocol Modules that translate them into Native Protocol Programming Interface(NPPI).
We provide a \texttt{DeviceModule} and \texttt{ProtocolModule} base classes, which have interface to Agent, receive commands from control applications and execute proper function implementation according to requested commands. 
In order to be used within the UniFlex framework, a device's NDPI has to be wrapped with Device Module that inherits from base class. The same applies to protocol, i.e. its NPPI has to be wrapped with class inheriting from Protocol Module class.
An example of function implementation is presented in Listing~\ref{list:ndi_to_upi_wrapper}.
Here, a \texttt{wifi\_set\_channel} function takes \emph{channel} as an argument and use NETLINK interface to communicate with Linux 802.11 subsystem to configure the network device. 
We provide \textit{bind\_function} decorator to mask function names which can also be used to implement an unified abstraction layer. In example, the function is hidden behind proper operation from UPI definition.
Note that the Device and Protocol Modules are Python objects and they keep state between consecutive function calls. 
Finally, as the Agent takes care about parsing of function arguments and serialization of return values, developer of Device/Protocol Module do not have to care about it. 

The UniFlex framework creates a \texttt{DeviceProxy} object for each Device Module and \texttt{ProtocolProxy} object for each Protocol Module present in controlled node and passes them to all control applications running on top. Such a proxy object contains all functions of device/protocol module and corresponding functions are bound together, i.e. calling function on proxy object translates in execution of function in device/protocol module object. This way, we achieve location-transparency. 

We manage to make Device and Protocol Modules very \emph{thin} and without dependencies to UniFlex core framework, so once implemented it may be used as part of the framework, but also as a standalone application.

\lstset{language=Python,
        basicstyle=\ttfamily\footnotesize,
        keywordstyle=\color{deepblue},
        commentstyle=\color{comments},
        stringstyle=\color{deepgreen},
				emphstyle=\ttb\color{deepred},    
        showstringspaces=false,
        procnamekeys={def,class}}

\begin{lstlisting}[caption=Example implementation of a device module function, label=list:ndi_to_upi_wrapper]

@bind_function(upi.radio.set_channel)
def wifi_set_channel(self, channel): 
    self.channel = channel 
    # set channel in wireless interface using NETLINK 
    # .......
    return reponse

\end{lstlisting}

As an example, we provide two device modules for different implementation of the same wireless technology, i.e. IEEE 802.11. The \texttt{WiFiAth9kDeviceModule} translates Python function calls into NDPI of COTS Atheros 9500 wireless card (mostly into NETLINK calls), while \\
\texttt{WiFiGnuRadioDeviceModule} maps them into parameter exposed by GNU-Radio engine. We have modified the 802.11 implementation~\cite{bloessl2013towards} in order to expose internal parameters of Wi-Fi protocol stack for control purpose.

%
%
\section{Deployment}

We adopted \emph{yaml} format for preparation and storing configuration for Agent. Such configuration file is loaded by Agent on it start and contains information about where to connect (Broker's URIs) and network applications as well as device modules to be started. The example configuration file is show on Listing~\ref{list:config_example}. In order to load control application or device module, one has to specify its source (source file or Python module) and give a name of the class. It is possible to pass dictionary of arguments to class constructor using \textit{kwargs} attribute. A Device Module is additionally given a name of device that it is serving. It is stored in \textit{device} object attribute.

\begin{lstlisting}[language=Python, caption=Example of Agent configuration file, label=list:config_example]
agent:
  name: 'LocalWiFiController'
  info: 'Distributed WiFi controller - local part'
  sub:  'tcp://192.168.1.1:8990' #Broker XPUB
  pub:  'tcp://192.168.1.1:8989' #Broker XSUB
  
# List of control apps
applications:
  local_wifi_controller:
      file : local_wifi_controller.py
      class_name : MyLocalWiFiController

# List of device modules
modules:
  wifi_ath9k:
      module : module_wifi_ath9k
      class_name : WifiModule
      device : 'phy0'
  wifi_gnuradio:
      module : module_wifi_gnuradio
      class_name : WifiModule
      device : 'uhd0'
      kwargs: {UsrpServer='http://10.0.0.1:8080'}
\end{lstlisting}

%
%
\section{Support of other Programming Languages}\label{sec:support_other}

Since the UniFlex prototype is implemented around ZMQ library, that is available for most of programming languages, support of other programming languages like C/C++ is possible. As a proof, we provide basic support for Node-RED (see Sect.~\ref{node-red-subsection}) and the Android OS.

%
%
\chapter{Integration with External Software Libraries}

There exists a wide range of specialized open-source software libraries and tools for data processing, mining, visualization, machine learning, etc. We argue that a control framework has to provide integration with such external tools in order to be widely used. This section gives a brief overview of currently supported integration with external software.

\section{Python Scientific Packages}

As UniFlex control applications are written using Python programming language, network developers can easily import any Python module within her network application. There exists number of scientific libraries for Python language including tools for data mining (SciPy), data processing (NumPy), machine learning (Tensorflow, PyBrain), etc.

%
%
\section{Node-RED Integration}\label{node-red-subsection}

Flow graphs are a great abstraction model with sufficient computational power to be able to program complex control behavior. Node-RED\footnote{\url{http://nodered.org/}} is a tool used by Internet-of-Things (IoT) community for wiring together hardware devices, APIs and online services in new and interesting ways. The UniFlex framework provides the possibility to connect to Node-RED. In a Node-RED flow graph, it is possible to subscribe to any event and also generate events, which are afterwards are processed by UniFlex control application(s).

\begin{figure}[!ht]
   \begin{center}
       \includegraphics[width=0.6\linewidth]{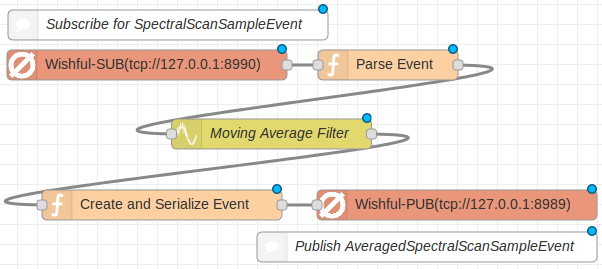}
   \end{center}
    \vspace{-10pt}
   \caption{Example Node-RED flowgraph consuming and producing UniFlex events.}
   \label{fig:node_red_example}
\end{figure}

Fig.~\ref{fig:node_red_example} shows an example Node-RED flowgraph illustrating how UniFlex events can be consumed and produced in Node-RED. It subscribes to \textit{SpectralScanSampleEvent}, calculates a moving average and publishes the result as an \textit{AverageSpectralScanSampleEvent} event, which can afterwards be processed by another UniFlex control application.

With usage of ZMQ implementation for node.js, we provide two new Node-RED nodes:
i) UniFlex Subscriber Node that connects to Broker and receives events of pre-configured type; and ii) UniFlex Publisher Node that sends events to Broker. 

%
%
\section{Mininet Integration}

In order to offer the application developer an easy way to test own network control programs, before deploying them in a real testbed, our framework can be executed in Mininet~\cite{mininet}, a container-based emulation which is able to emulate large network topologies on a single computer. Specifically, we use Mininet-WiFi~\cite{mininetwifigit},~\cite{fontes2015mininet} which allows rapid prototyping and experimental evaluation of control programs for wireless environments by augmenting the well-known Mininet emulator with virtual 802.11 WiFi stations and access points. Hence, it allows the emulation of control programs requiring access to the higher 802.11 MAC protocol stack, aka SoftMAC~\cite{mac80211}.



\chapter{Applications}

In this section we present selected network apps we implemented using UniFlex framework.

\section{Wireless Topology Monitor}

Knowledge of wireless topology is of great importance when performing experiments within wireless testbeds as it provides the network programmer with a full view of the network. Usually, optimization of wireless networks is touching configuration of neighboring nodes (channel assignment, load balancing, station handover, D2D, etc).

We implemented topology monitoring as a control application in UniFlex. It consists of local application running on each wireless node that collects and aggregates information about wireless neighbors and sends it up to \textit{Wireless Topology Monitor} application residing in central compute node. For example, in case of 802.11 infrastructure networks the local app running on each AP would periodically report the set of associated client stations. 

In general the information about all wireless neighbors, a so-called hearing map, is desired, which can be obtained by the local app by means of packet sniffing over a monitor interface, i.e. from all received frames the packet headers are processed to get information about transmitter address as well as additional information like signal strength and bitrate. For efficiency reasons the local apps do not send the sniffed frames directly to the \textit{Wireless Topology Monitor} app, instead only aggregated information like transmitter address, average signal strength and bitrate are send via events to central \textit{Wireless Topology Monitor} app.

Based on the received information from the local network apps the \textit{Wireless Topology Monitor} app is able to create the wireless topology graph. Note, that even within a single technology a topology may be defined in multiple ways. It may represent so called hearing map, where an edge is between nodes in communication range, or a connectivity map. While a hearing map may be used to discover handover or D2D opportunities, a connectivity map may be used to control link transmission parameters.

The \textit{Topology Monitor} app may be parameterized to generate topology graph of required type(s). Moreover, its interfaces are technology independent, i.e. being given properly structured and described input data, it is able to generate topology graph.

The nodes in the topology graph contain unique identifiers (UUID) of UniFlex-enabled nodes, or transmitter MAC address for other nodes. Edges represent connections between nodes and their weights take value of pre-configured parameter, like bitrate, signal strength, etc. The topology graph contains also meta-data describing it. In multi-channel environments the topology graph is a set of graphs, one for each radio channel.

Any other network app can subscribe to events generated by \textit{Topology Monitor} app, e.g. notification about topology changes. Note, the network nodes in topology graph can be directly accessed as UniFlex provides helper functions mapping node identifiers to node reference objects (if known).

%
%
\section{Mobility Management}
\begin{figure}[!ht]
   \begin{center}
       \includegraphics[width=0.6\linewidth]{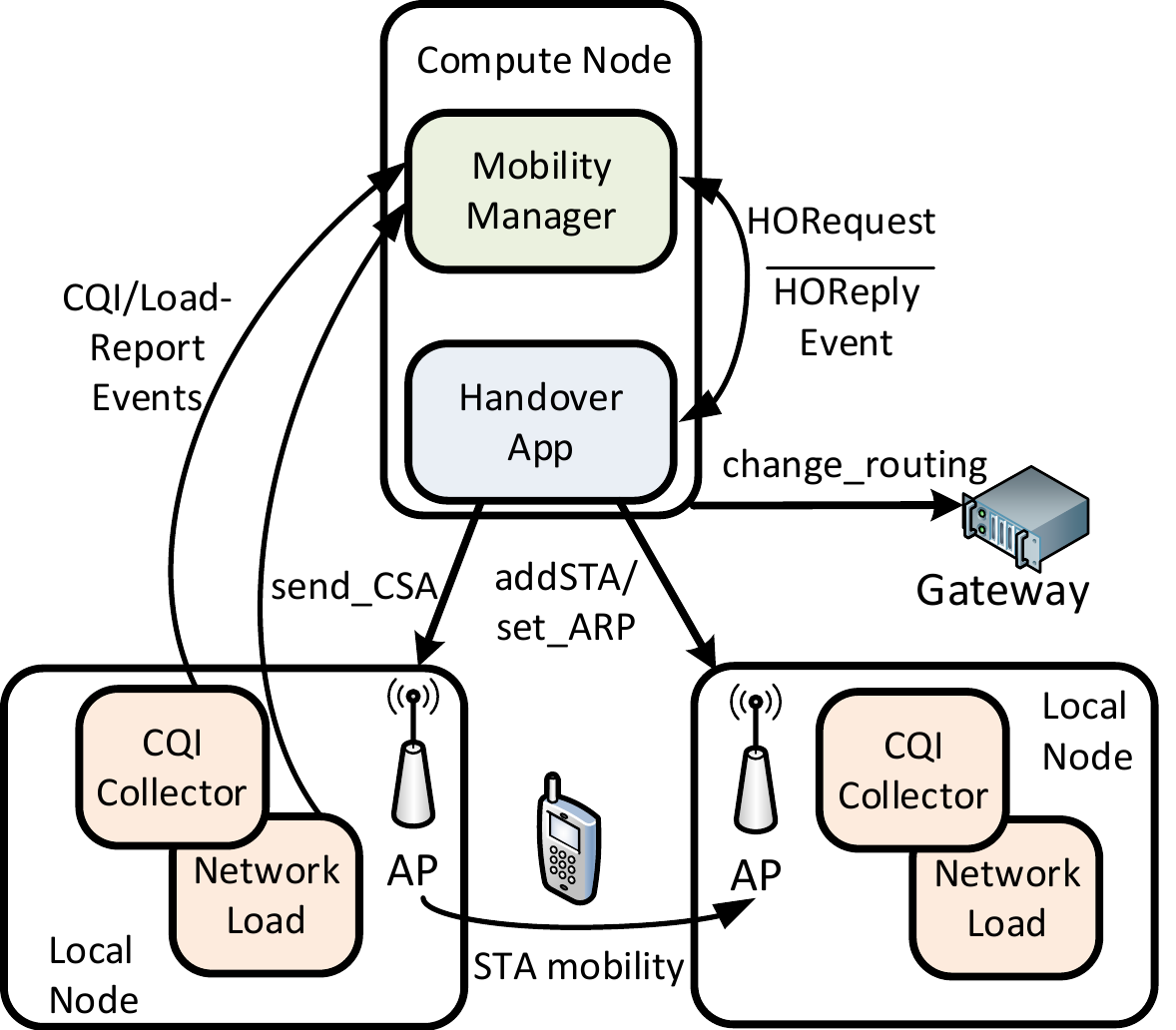}
   \end{center}
    \vspace{-10pt}
   \caption{Mobility management for enterprise 802.11 networks.}
   \label{fig:ho_app}
\end{figure}
Novel applications, e.g. mobile HD video, and devices, e.g. smartphones and tablets, require much better mobility support and higher QoS/QoE. Therefore, in~\cite{Zubow16bigap_seamless_handover,bigap2} we presented BIGAP, a seamless handover scheme for high performance enterprise IEEE 802.11 networks. We implemented the mobility management function of BIGAP in UniFlex. Fig.~\ref{fig:ho_app} shows the hierarchical controller architecture consisting of a central network app and multiple local network apps. On each AP two local control applications are running for collection of information, i.e. quality of the active wireless links as well as potential links to client stations in communication range which are currently being served by another co-located AP (\textit{CQIReportEvent}). This data as well as information about the current network load at each AP (\textit{LoadReportEvent}) is reported as event to the central mobility manager network app which decides on handover by sending out a \textit{HORequest}. The latter is processed by the \textit{Handover App} which performs the actual handover operation as described in~\cite{Zubow16bigap_seamless_handover}.

%
%
\section{Interference Management}
\begin{figure}[!ht]
   \begin{center}
       \includegraphics[width=0.7\linewidth]{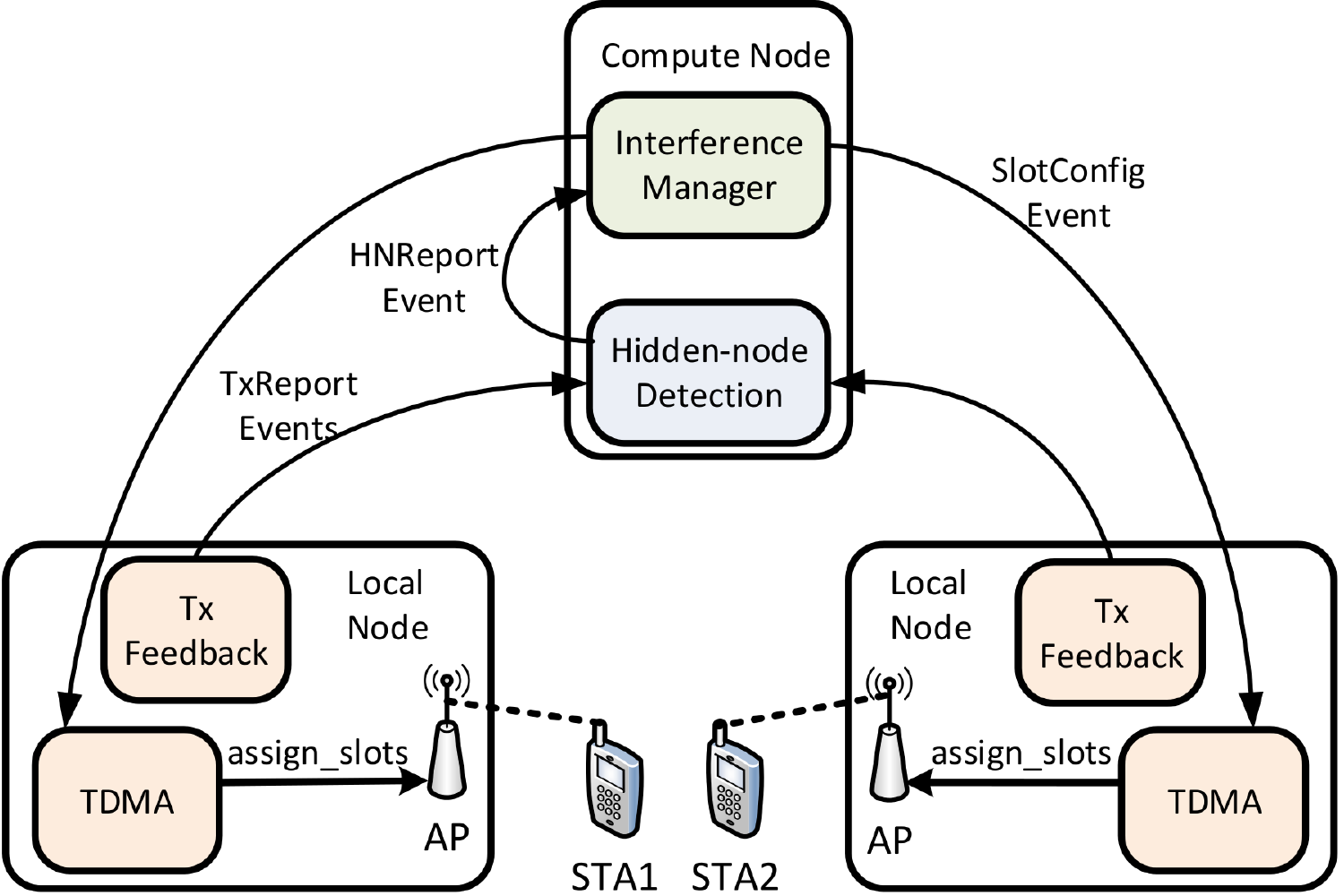}
   \end{center}
    \vspace{-10pt}
   \caption{Interference management through airtime management in 802.11 networks.}
   \label{fig:im_app}
\end{figure}
Another known problem experienced in 802.11 networks is performance degradation due to co-channel interference because of hidden nodes. The impact can be mitigated by preventing overlapping transmissions (in time) of affected nodes, e.g. APs, by efficient airtime management through interference avoidance techniques at the MAC layer. We implemented interference management in UniFlex using hMAC~\cite{zehlhmac}. In particular the following two features have been implemented: \textit{i)} detection of wireless links suffering from hidden nodes and \textit{ii)} execution of airtime management in which two wireless links suffering from the hidden node problem are getting exclusive time slots assigned.

Fig.~\ref{fig:im_app} illustrates the developed hierarchical controller architecture. Here we have two control applications running on each AP locally due to timing constraints and efficiency reasons. The \textit{TxFeedback} control application provides transmission feedback information like number of ARQ retries to the central \textit{Hidden-node detection} control application which is using this information for discovery of hidden-nodes. Each pair of wireless links suffering from hidden-node is reported using \textit{HNReportEvent} and consumed by another central control application, \textit{Interference Manager}, which in turn decides on the time slot configuration to be used. The actual assignment of time slots to nodes is performed by the local \textit{TDMA} scheduler.

%
%
\section{Device-to-Device}
\begin{figure}[!ht]
   \begin{center}
       \includegraphics[width=0.6\linewidth]{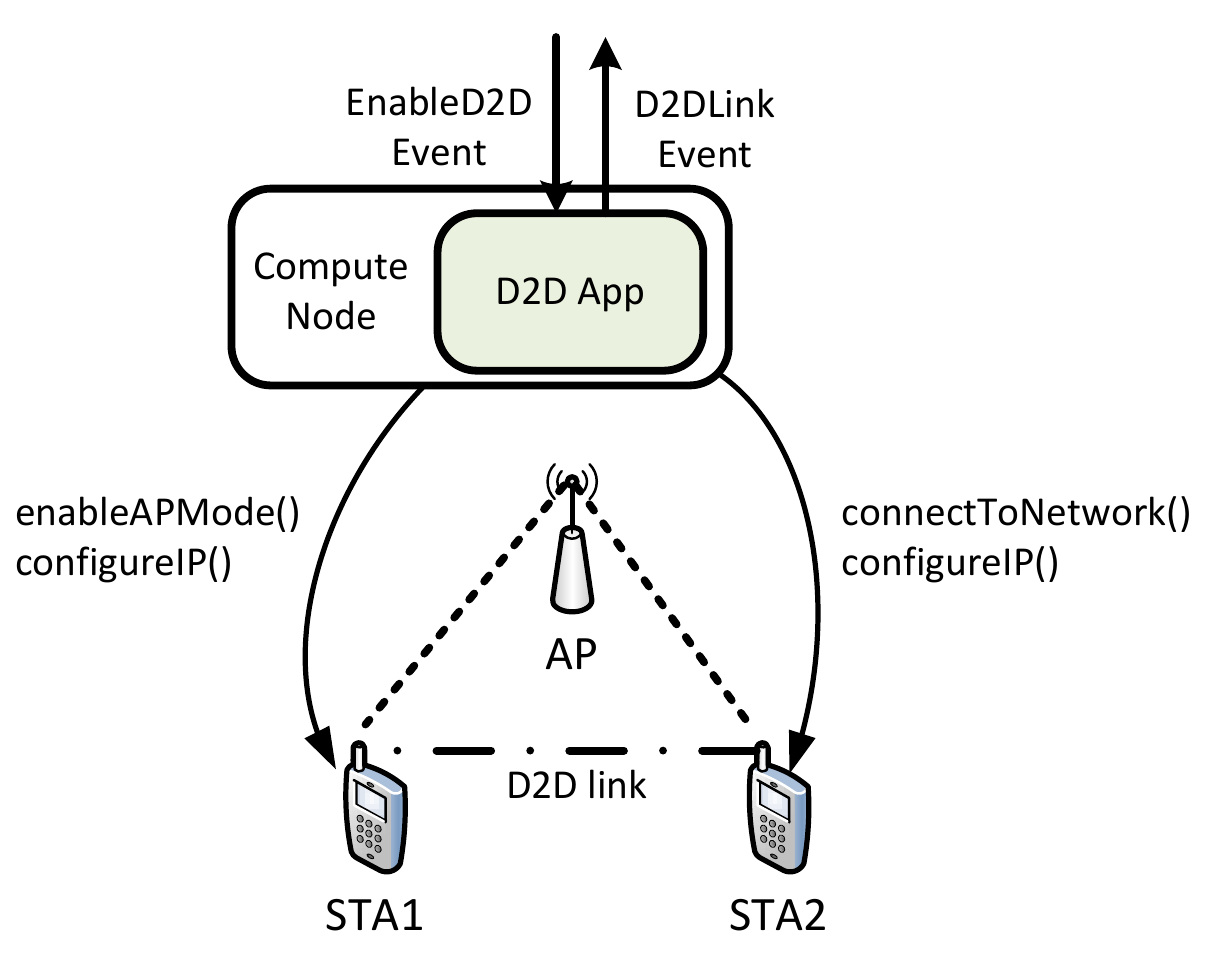}
   \end{center}
    \vspace{-10pt}
   \caption{UniFlex network app enabling direct device-to-device connectivity in IEEE 802.11 networks.}
   \label{fig:nfunc_d2d_example}
\end{figure}
The efficiency of wireless infrastructure networks can be significantly improved by enabling direct device-to-device (D2D) communication between stations (Fig.~\ref{fig:nfunc_d2d_example}). We implemented D2D functionality for 802.11 network as a central network app whose functionality can be used by other control programs. The interaction takes place by exchange of events, namely, \textit{EnableD2D} and \textit{D2DLink} event respectively. On receiving an \textit{EnableD2D} event the D2D network app performs remote API function calls on the network devices of the corresponding two client stations. On one station is enables soft AP mode whereas on the other station connects to that AP. Moreover, on both clients IP connectivity is configured using the appropriate API function calls. Note, all the implementation details are hidden from the high level control design, thus simplify a controller implementation.


\chapter{Evaluation}

In this section we analyze the performance of our prototypical implementation with respect to two categories: i) basic network operation and ii) scalability with respect to the number of controlled network nodes.

\section{Basic Network Operation}

Observing and modifying the network state by means of executing API functions is a basic building block of UniFlex operations, its performance is of great importance on the overall system’s performance. We identified latency for network state monitoring and API function execution as an important performance metric.

For this measurement, the experiments were conducted using three different network nodes: i) high performance Intel i7-4790, ii) small-form-factor-PC based on Intel NUC and iii) low-power single-board ARM Cortex-A8 machines (BeagleBone). All three nodes were equipped with a single 802.11 network device. For the evaluation of the performance of local calls we implemented a local control application whereas for remote calls a global controller running on a different node connected by Gigabit-Ethernet was used. We measured the latency of executing API functions, both locally and remotely. 

Table~\ref{tbleval1} shows the median (mean) and 99th percentile of the latency when executing a single blocking local API function call, \textit{get\_interfaces()} which returns the available wireless interfaces of a wireless node.

\begin{table}[ht!]
\centering
\begin{tabular}{@{} *5l @{}}    \toprule
\emph{Latency} & \emph{Median} & \emph{99 \%ile}  \\\midrule
Intel (i7-4790, 3.6\,GHz) & 0.4017\,ms  & 0.5009\,ms    \\ 
Intel NUC (i5-4250U, 1.3\,GHz) & 0.7627\,ms & 1.3986\,ms\\
BeagleBone (ARM armv7l, 1\,GHz) & 10.0138\,ms & 11.4258\,ms\\\bottomrule
\hline
\end{tabular}
\vspace{5pt}
\caption{Latency for executing single blocking local API function.}\label{tbleval1}
\end{table}

Further, Table~\ref{tbleval1_remote} shows the results when executing the same function remotely. Note that the network overhead for the execution of this API call is around 1600\,Bytes per call.

From the results we can conclude that the latency of performing an API call, locally or remotely, is sufficient low to be used for real-world control applications. However, when using slow ARM SoCs the latency is $11-25\times$ larger as compared to i7-4790 which might be insufficient. However, we argue that the UniFlex agent can be easily implemented in a low-level programming language like C.

\begin{table}[h]
\centering
\begin{tabular}{@{} *5l @{}}    \toprule
\emph{Latency} & \emph{Median} & \emph{99 \%ile}  \\\midrule
Intel (i7-4790, 3.6\,GHz) & 1.2896\,ms & 1.5042\,ms\\
Intel NUC (i5-4250U, 1.3\,GHz) & 2.6748\,ms & 3.1662\,ms\\
BeagleBone (ARM armv7l, 1\,GHz) & 14.5829\,ms & 16.4588\,ms\\\bottomrule
\hline
\end{tabular}
\vspace{5pt}
\caption{Latency for executing single blocking remote API function.}\label{tbleval1_remote}
\end{table}

\section{Scalability}

Another important performance metric is scalability. A key feature of our framework is its distributed architecture for scale-out performance. As the number of network nodes to be controlled grows the demand on the control plane increases. 

For this measurement, the experiments were conducted in the ORBIT testbed~\cite{raychaudhuri2005overview} consisting of i7-4790 x86 machines. The number of controlled network nodes was varied from one to 87 nodes. A single central control program was executing API calls, \textit{get\_interfaces()}, on each node using non-blocking calling semantic. We measured the latency to get the results from all nodes.

\begin{figure}[!ht]
   \begin{center}
       \includegraphics[width=0.8\linewidth]{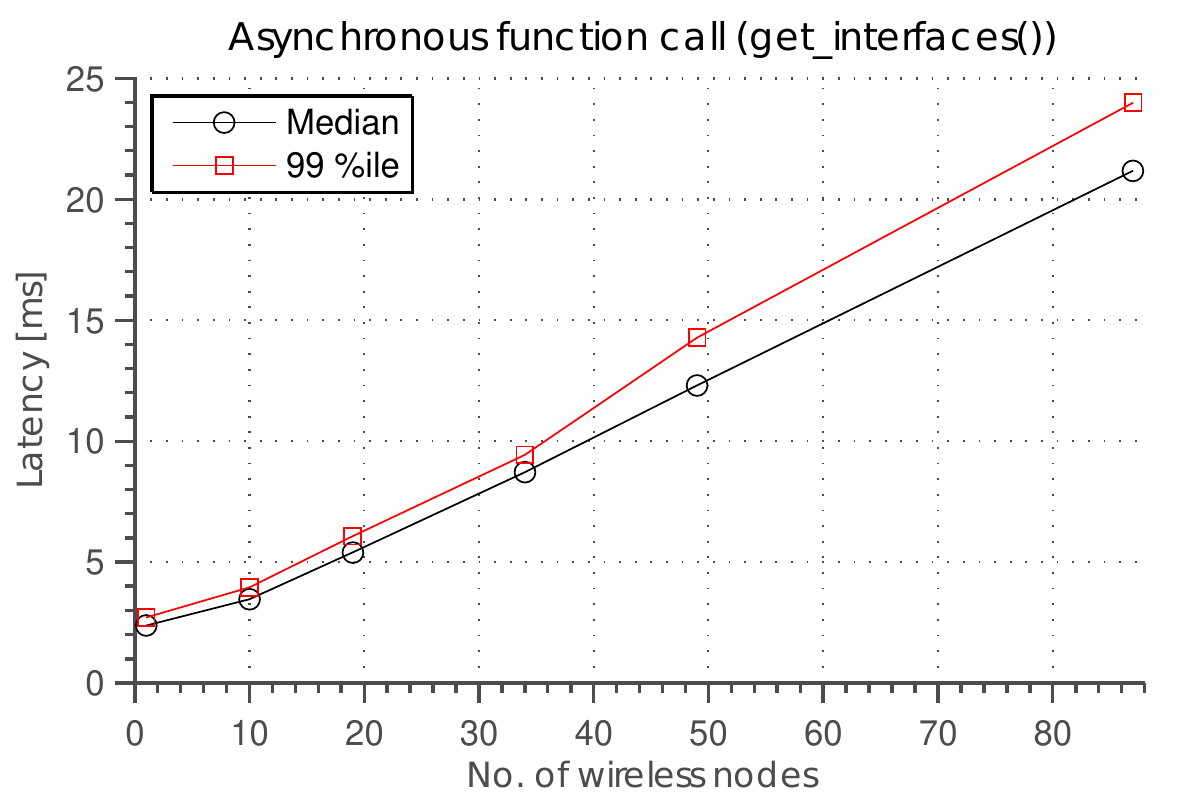}
   \end{center}
    \vspace{-10pt}
   \caption{Latency for executing single non-blocking API function call on a set of nodes.}
   \label{fig:latency}
\end{figure}

The results are shown in Fig.~\ref{fig:latency}. It takes less than 25\,ms to execute a non-blocking API call on all 87 network nodes. Note, that the latency per API call decreases with the number of nodes, i.e. 2.37\,ms vs. 0.24\,ms for 1 and 87 nodes respectively. This is because non-blocking calls are executed in parallel.

Note, that with 87 nodes and a API calling rate of 10\,Hz the control plane workload at the central controller is already high, i.e. 16\,Mbit/s. In order to reduce it the use of hierarchical or local controllers is advisable.

\chapter{Related Work}

\noindent
Related work falls into three categories:

\medskip

\textbf{Cross-layer Control:} CRAWLER~\cite{aktacs2014framework,aktas2012crawler} is experimentation architecture for centralized network monitoring and cross-layer coordination over different devices. ClickWatch~\cite{scheidgen2012clickwatch} aims for simplification of experimentation of wireless cross-layer solutions implemented using the Click Modular Router~\cite{kohler2000click}. Both frameworks aim to facilitate experimentation and offer possibility to control all nodes in the network from a single centralized controller.
In contrast, UniFlex is more flexible as it allows distributing controller logic over multiple nodes so that time sensitive control logic can be executed directly on the network node to be controlled.

\medskip

\textbf{Software-defined Networking:} There are already lots of distributed control frameworks, but they are mostly focused on control of wired switches using open protocols (e.g. OpenFlow). Some of them, like ONOS~\cite{berde2014onos} and ONIX~\cite{onix} are focused on scalability and performance. As they are already in very advanced state, it is hard to use them for resource constrained devices or to adjust them to wireless networking. Ryuo~\cite{ryuo} and Kandoo~\cite{hassas2012kandoo} provide the possibility for offloading of control applications to local controllers as a way to limit the control plane load. Local controllers handle frequent events, while a logically centralized root controller handles rare events. In contrast UniFlex is not restricted to two levels of controllers as it allows direct communication between any controller applications. Beehive~\cite{beehive1}\cite{beehive2} provides interesting features like automatic distribution of network applications over network nodes. While having similar concepts, Beehive does not differentiate between control applications and device modules which are of great importance when targeting the control of heterogeneous wireless networks.

CoAP~\cite{patro2015coap} proposes a vendor neutral centralized framework for configuration, coordination and management of residential 802.11 APs using an open API implemented over  OpenFlow~\cite{mckeown2008openflow}. In contrast to UniFlex the CoAP API is restricted to control of 802.11 networks. Moreover, only centralized control programs are possible. OpenRF~\cite{kumar2013bringing} provides programming abstractions tailored for wireless networks, i.e. MIMO interference management techniques that impact the physical layer. OpenRF is restricted to centralized control of 802.11 infrastructure networks. Finally, in \cite{zubow2015sdn} SDN architecture for centralized spectrum brokerage in residential infrastructure Cognitive Radio networks was proposed.


\medskip

\textbf{General Distributed Control Platforms:} ROS~\cite{quigley2009ros} is an open source robot operating system for rapid prototyping. ROS is focused on providing control for a single robot, trying to achieve one goal, and having all devices working towards that goal. In UniFlex, we are trying to achieve harmonization of multiple devices. Moreover, we also provide time scheduled execution of operations on multiple devices.


\chapter{Conclusions}

This paper introduces UniFlex, a framework that uses SDN concepts to simplify prototyping of novel wireless networking solutions requiring cross-layer control coordinated among multiple heterogeneous wireless network nodes. It provides a rich API for management and control of operation of network devices. 
Control applications representing controller logic can be implemented in a local, central or distributed manner. This allows to place time-sensitive control functions close to controlled device, off-load resource hungry application to compute servers and make them work together to control the entire network.


\begingroup
\renewcommand{\cleardoublepage}{}
\renewcommand{\clearpage}{}
\chapter{Acknowledgments}\label{chap:ack}
\endgroup

This work has been supported by the European Union’s Horizon 2020 research and innovation programme under grant agreement No. 645274 (WiSHFUL project).

\bibliographystyle{IEEEtran}
\bibliography{biblio,IEEEabrv}

\begin{thebibliography}{10}
\providecommand{\url}[1]{#1}
\csname url@samestyle\endcsname
\providecommand{\newblock}{\relax}
\providecommand{\bibinfo}[2]{#2}
\providecommand{\BIBentrySTDinterwordspacing}{\spaceskip=0pt\relax}
\providecommand{\BIBentryALTinterwordstretchfactor}{4}
\providecommand{\BIBentryALTinterwordspacing}{\spaceskip=\fontdimen2\font plus
\BIBentryALTinterwordstretchfactor\fontdimen3\font minus
  \fontdimen4\font\relax}
\providecommand{\BIBforeignlanguage}[2]{{%
\expandafter\ifx\csname l@#1\endcsname\relax
\typeout{** WARNING: IEEEtran.bst: No hyphenation pattern has been}%
\typeout{** loaded for the language `#1'. Using the pattern for}%
\typeout{** the default language instead.}%
\else
\language=\csname l@#1\endcsname
\fi
#2}}
\providecommand{\BIBdecl}{\relax}
\BIBdecl

\bibitem{mvulla2015analysis}
J.~Mvulla, E.-C. Park, M.~Adnan, and J.-H. Son, ``{Analysis of asymmetric
  hidden node problem in IEEE 802.11 ax heterogeneous WLANs},'' in
  \emph{{Information and Communication Technology Convergence (ICTC), 2015
  International Conference on}}.\hskip 1em plus 0.5em minus 0.4em\relax IEEE,
  2015, pp. 539--544.

\bibitem{pejovic2014whiterate}
V.~Pejovic and E.~M. Belding, ``{Whiterate: A context-aware approach to
  wireless rate adaptation},'' \emph{IEEE Transactions on Mobile Computing},
  vol.~13, no.~4, pp. 921--934, 2014.

\bibitem{kumar2013bringing}
S.~Kumar, D.~Cifuentes, S.~Gollakota, and D.~Katabi, ``{Bringing cross-layer
  MIMO to today's wireless LANs},'' in \emph{{ACM SIGCOMM Computer
  Communication Review}}, vol.~43, no.~4.\hskip 1em plus 0.5em minus
  0.4em\relax ACM, 2013, pp. 387--398.

\bibitem{ryuo}
S.~Zhang, Y.~Shen, M.~Herlich, K.~Nguyen, Y.~Ji, and S.~Yamada, ``{Ryuo: Using
  high level northbound API for control messages in software defined
  network},'' in \emph{{Network Operations and Management Symposium (APNOMS),
  2015 17th Asia-Pacific}}, Aug 2015, pp. 115--120.

\bibitem{SAI}
\BIBentryALTinterwordspacing
``{Switch Abstraction Interface (SAI) as part of Open Compute project}.''
  [Online]. Available:
  \url{https://github.com/opencomputeproject/OCP-Networking-Project-Community-Contributions}
\BIBentrySTDinterwordspacing

\bibitem{ruckebusch2016unified}
P.~Ruckebusch, S.~Giannoulis, E.~De~Poorter, I.~Moerman, I.~Tinnirello,
  D.~Garlisi, P.~Gallo, N.~Kaminski, L.~DaSilva, P.~Gawlowicz \emph{et~al.},
  ``{A unified radio control architecture for prototyping adaptive wireless
  protocols},'' in \emph{{Networks and Communications (EuCNC), 2016 European
  Conference on}}.\hskip 1em plus 0.5em minus 0.4em\relax IEEE, 2016, pp.
  58--63.

\bibitem{fortuna2015wireless}
C.~Fortuna, P.~Ruckebusch, C.~Van~Praet, I.~Moerman, N.~Kaminski, L.~DaSilva,
  I.~Tinirello, G.~Bianchi, F.~Gringoli, A.~Zubow \emph{et~al.}, ``{Wireless
  software and hardware platforms for flexible and unified radio and network
  control},'' in \emph{{European Conference on Networks and Communications
  (Eu-CNC)}}, 2015.

\bibitem{zeromq-2014}
iMatix Corporation, ``{ZMQ - Code Connected},'' \textit{http://zeromq.org/},
  January 2014, accessed: 2015-08-04.

\bibitem{bloessl2013towards}
B.~Bloessl, M.~Segata, C.~Sommer, and F.~Dressler, ``{Towards an Open Source
  IEEE 802.11p Stack: A Full SDR-based Transceiver in GNURadio},'' in \emph{5th
  IEEE Vehicular Networking Conference (VNC 2013)}.\hskip 1em plus 0.5em minus
  0.4em\relax Boston, MA: IEEE, December 2013, pp. 143--149.

\bibitem{mininet}
\BIBentryALTinterwordspacing
``{Mininet}.'' [Online]. Available: \url{http://mininet.org/}
\BIBentrySTDinterwordspacing

\bibitem{mininetwifigit}
\BIBentryALTinterwordspacing
``{Mininet-WiFi}.'' [Online]. Available:
  \url{https://github.com/intrig-unicamp/mininet-wifi}
\BIBentrySTDinterwordspacing

\bibitem{fontes2015mininet}
R.~R. Fontes, S.~Afzal, S.~H. Brito, M.~A. Santos, and C.~E. Rothenberg,
  ``{Mininet-WiFi: Emulating software-defined wireless networks},'' in
  \emph{{Network and Service Management (CNSM), 2015 11th International
  Conferenc}e on}.\hskip 1em plus 0.5em minus 0.4em\relax IEEE, 2015, pp.
  384--389.

\bibitem{mac80211}
\BIBentryALTinterwordspacing
``{Linux wireless - mac80211}.'' [Online]. Available:
  \url{https://wireless.wiki.kernel.org/en/developers/documentation/mac80211}
\BIBentrySTDinterwordspacing

\bibitem{Zubow16bigap_seamless_handover}
A.~Zubow, S.~Zehl, and A.~Wolisz, ``{BIG AP -- Seamless Handover in High
  Performance Enterprise IEEE 802.11 Networks},'' in \emph{{Network Operations
  and Management Symposium (NOMS), 2016 IEEE}}, April 2016.

\bibitem{bigap2}
S.~Zehl, A.~Zubow, and A.~Wolisz, ``Bigap - a seamless handover scheme for high
  performance enterprise ieee 802.11 networks,'' in \emph{NOMS 2016 - 2016
  IEEE/IFIP Network Operations and Management Symposium}, April 2016, pp.
  1015--1016.

\bibitem{zehlhmac}
------, ``{hMAC: Enabling Hybrid TDMA/CSMA on IEEE 802.11 Hardware},''
  Telecommunication Networks Group, Technische Universit\"at Berlin, Tech. Rep.
  TKN-16-004, November 2016.

\bibitem{raychaudhuri2005overview}
D.~Raychaudhuri, I.~Seskar, M.~Ott, S.~Ganu, K.~Ramachandran, H.~Kremo,
  R.~Siracusa, H.~Liu, and M.~Singh, ``{Overview of the ORBIT radio grid
  testbed for evaluation of next-generation wireless network protocols},'' in
  \emph{{IEEE Wireless Communications and Networking Conference, 2005}},
  vol.~3.\hskip 1em plus 0.5em minus 0.4em\relax IEEE, 2005, pp. 1664--1669.

\bibitem{aktacs2014framework}
I.~Akta{\c{s}}, O.~Pun{\~n}al, F.~Schmidt, T.~Dr{\"u}ner, and K.~Wehrle, ``{A
  framework for remote automation, configuration, and monitoring of real-world
  experiments},'' in \emph{{Proceedings of the 9th ACM international workshop
  on Wireless network testbeds, experimental evaluation and
  characterization}}.\hskip 1em plus 0.5em minus 0.4em\relax ACM, 2014, pp.
  9--16.

\bibitem{aktas2012crawler}
I.~Aktas, F.~Schmidt, M.~H. Alizai, T.~Dr{\"u}ner, and K.~Wehrle, ``{CRAWLER:
  An experimentation platform for system monitoring and
  cross-layer-coordination},'' in \emph{{World of Wireless, Mobile and
  Multimedia Networks (WoWMoM), 2012 IEEE International Symposium on a}}.\hskip
  1em plus 0.5em minus 0.4em\relax IEEE, 2012, pp. 1--9.

\bibitem{scheidgen2012clickwatch}
M.~Scheidgen, A.~Zubow, and R.~Sombrutzki, ``Clickwatch—an experimentation
  framework for communication network test-beds,'' in \emph{2012 IEEE Wireless
  Communications and Networking Conference (WCNC)}.\hskip 1em plus 0.5em minus
  0.4em\relax IEEE, 2012, pp. 3296--3301.

\bibitem{kohler2000click}
E.~Kohler, R.~Morris, B.~Chen, J.~Jannotti, and M.~F. Kaashoek, ``The click
  modular router,'' \emph{ACM Transactions on Computer Systems (TOCS)},
  vol.~18, no.~3, pp. 263--297, 2000.

\bibitem{berde2014onos}
P.~Berde, M.~Gerola, J.~Hart, Y.~Higuchi, M.~Kobayashi, T.~Koide, B.~Lantz,
  B.~O'Connor, P.~Radoslavov, W.~Snow \emph{et~al.}, ``{ONOS: towards an open,
  distributed SDN OS},'' in \emph{{Proceedings of the third workshop on Hot
  topics in software defined networking}}.\hskip 1em plus 0.5em minus
  0.4em\relax ACM, 2014, pp. 1--6.

\bibitem{onix}
\BIBentryALTinterwordspacing
T.~Koponen, M.~Casado, N.~Gude, J.~Stribling, L.~Poutievski, M.~Zhu,
  R.~Ramanathan, Y.~Iwata, H.~Inoue, T.~Hama, and S.~Shenker, ``Onix: A
  distributed control platform for large-scale production networks,'' in
  \emph{Proceedings of the 9th USENIX Conference on Operating Systems Design
  and Implementation}, ser. OSDI'10.\hskip 1em plus 0.5em minus 0.4em\relax
  Berkeley, CA, USA: USENIX Association, 2010, pp. 351--364. [Online].
  Available: \url{http://dl.acm.org/citation.cfm?id=1924943.1924968}
\BIBentrySTDinterwordspacing

\bibitem{hassas2012kandoo}
S.~Hassas~Yeganeh and Y.~Ganjali, ``{Kandoo: a framework for efficient and
  scalable offloading of control applications},'' in \emph{{Proceedings of the
  first workshop on Hot topics in software defined networks}}.\hskip 1em plus
  0.5em minus 0.4em\relax ACM, 2012, pp. 19--24.

\bibitem{beehive1}
\BIBentryALTinterwordspacing
S.~H. Yeganeh and Y.~Ganjali, ``Beehive: Towards a simple abstraction for
  scalable software-defined networking,'' in \emph{Proceedings of the 13th ACM
  Workshop on Hot Topics in Networks}, ser. HotNets-XIII.\hskip 1em plus 0.5em
  minus 0.4em\relax New York, NY, USA: ACM, 2014, pp. 13:1--13:7. [Online].
  Available: \url{http://doi.acm.org/10.1145/2670518.2673864}
\BIBentrySTDinterwordspacing

\bibitem{beehive2}
\BIBentryALTinterwordspacing
------, ``Beehive: Simple distributed programming in software-defined
  networks,'' in \emph{Proceedings of the Symposium on SDN Research}, ser. SOSR
  '16.\hskip 1em plus 0.5em minus 0.4em\relax New York, NY, USA: ACM, 2016, pp.
  4:1--4:12. [Online]. Available:
  \url{http://doi.acm.org/10.1145/2890955.2890958}
\BIBentrySTDinterwordspacing

\bibitem{patro2015coap}
A.~Patro and S.~Banerjee, ``{COAP: A software-defined approach for home WLAN
  management through an open API},'' \emph{ACM SIGMOBILE Mobile Computing and
  Communications Review}, vol.~18, no.~3, pp. 32--40, 2015.

\bibitem{mckeown2008openflow}
N.~McKeown, T.~Anderson, H.~Balakrishnan, G.~Parulkar, L.~Peterson, J.~Rexford,
  S.~Shenker, and J.~Turner, ``{OpenFlow: enabling innovation in campus
  networks},'' \emph{ACM SIGCOMM Computer Communication Review}, vol.~38,
  no.~2, pp. 69--74, 2008.

\bibitem{zubow2015sdn}
A.~Zubow, M.~D{\"o}ring, M.~Chwalisz, and A.~Wolisz, ``{A SDN approach to
  spectrum brokerage in infrastructure-based Cognitive Radio networks},'' in
  \emph{{Dynamic Spectrum Access Networks (DySPAN), 2015 IEEE International
  Symposium on}}.\hskip 1em plus 0.5em minus 0.4em\relax IEEE, 2015, pp.
  375--384.

\bibitem{quigley2009ros}
M.~Quigley, K.~Conley, B.~Gerkey, J.~Faust, T.~Foote, J.~Leibs, R.~Wheeler, and
  A.~Y. Ng, ``{ROS: an open-source Robot Operating System},'' in \emph{{ICRA
  workshop on open source software}}, vol.~3, no. 3.2.\hskip 1em plus 0.5em
  minus 0.4em\relax Kobe, Japan, 2009, p.~5.

\end{thebibliography}


\chapter{Appendix A -- Description of UniFlex APIs}\label{sec:appendix_a}

%
%
\begin{table}[!ht]
\centering
\caption{Agent API}
\label{api:agent}
\begin{tabularx}{\textwidth}{@{}XX@{}} \toprule
Function                          & Description     \\ 
\hline

Boolean::add\_control\_application(\emph{controlAppObj}) & Add ControlApplication object to Agent.\\ 
 & Returns \emph{True} if succeeded; otherwise \emph{False}\\
\hline

Boolean::remove\_control\_application(\emph{uuid}) & Remove ControlApplication object from Agent by its UUID.\\ 
 & Returns \emph{True} if succeeded; otherwise \emph{False}\\
\hline

ControlApplicationList::get\_control\_applications() & Get all Control Application of Agent.\\ 
 & Returns list of ControlApplication objects. \\
\hline

ControlApplication::get\_control\_application(\emph{uuid}) & Get ControlApplication object by its UUID.\\ 
 & Returns Control Application object. \\
\hline

Boolean::start\_control\_application(\emph{uuid}) & Start Control Application by its UUID.\\ 
 & Returns \emph{True} if succeeded; otherwise \emph{False}\\
\hline

Boolean::stop\_control\_application(\emph{uuid}) & Stop Control Application by its UUID.\\ 
 & Returns \emph{True} if succeeded; otherwise \emph{False}\\
\hline

Boolean::add\_device\_module(\emph{deviceModuleObj}) & Add DeviceModule object to Agent.\\ 
 & Returns \emph{True} if succeeded; otherwise \emph{False}\\
\hline

DeviceModuleList::get\_device\_modules() & Get all Device Modules installed in Agent.\\ 
 & Returns list of DeviceModule objects. \\
\hline

DeviceModule::get\_device\_module(\emph{uuid}) & Get DeviceModule object by its UUID.\\ 
 & Returns DeviceModule object. \\
\hline

Boolean::add\_protocol\_module(\emph{protModuleObj}) & Add ProtocolModule object to Agent.\\ 
 & Returns \emph{True} if succeeded; otherwise \emph{False}\\
\hline

ProtocolModuleList::get\_protocol\_modules() & Get all Protocol Modules installed in Agent.\\ 
 & Returns list of ProtocolModule objects. \\
\hline

ProtocolModule::get\_protocol\_module(\emph{uuid}) & Get ProtocolModule object by its UUID.\\ 
 & Returns ProtocolModule object. \\
\hline

\end{tabularx}
\end{table}

%
%
\begin{table}[!ht]
\centering
\caption{Module API}
\label{api:module}
\begin{tabularx}{\textwidth}{@{}XX@{}} \toprule
Function                          & Description     \\ \hline

Boolean::send\_event(\emph{event}, \emph{mode}) & Sent event using one of two modes: \emph{node-broadcast} and \emph{global-broadcast}.\\ 
 & Returns \emph{True} if succeeded; otherwise \emph{False}\\
\hline

\end{tabularx}
\end{table}

%
%
\begin{table}[!ht]
\centering
\caption{Device Module API}
\label{api:device_module}
\begin{tabularx}{\textwidth}{@{}XX@{}} \toprule
Function                          & Description     \\ \hline
operation\_1(\emph{arg1,...}) & Python wrapper for operation exposed by device.\\ \hline
operation\_N(\emph{arg1,...}) & Python wrapper for operation exposed by device. \\ \hline
\end{tabularx}
\end{table}

%
%
\begin{table}[!ht]
\centering
\caption{Protocol Module API}
\label{api:protocol_module}
\begin{tabularx}{\textwidth}{@{}XX@{}} \toprule
Function                          & Description     \\ \hline
operation\_1(\emph{arg1,...}) & Python wrapper for operation exposed by protocol.\\ \hline
operation\_N(\emph{arg1,...}) & Python wrapper for operation exposed by protocol. \\ \hline
\end{tabularx}
\end{table}

%
%
\begin{table}[!ht]
\centering
\caption{Control Application API}
\label{api:control_app}
\begin{tabularx}{\textwidth}{@{}XX@{}} \toprule
Function                          & Description     \\ 
\hline

NodeProxy::get\_local\_node() & Get NodeProxy object for local node , i.e. the one that runs Application.\\ 
 & Returns NodeProxy object.\\
\hline

Boolean::send\_event(\emph{event}, \emph{mode}) & Sent event using one of two modes: \emph{node-broadcast} or \emph{global-broadcast}.\\
 & Returns \emph{True} if succeeded; otherwise \emph{False}\\
\hline

Boolean::subscribe\_for\_events(\emph{eventType}, \emph{callback}, \emph{mode}) & Subscribe for events of specific type using one of two modes:\\
 & - \emph{node-broadcast} –- subscribe for events of specific type generated on local node \\
 & - \emph{global-broadcast} -- subscribe for events of specific type generated at any node in network \\
 & The \emph{callback} function will be called on reception of event. \\
 & Note: If event type is not specified, application subscribes for events of all types.\\
 & Returns \emph{True} if succeeded; otherwise \emph{False}\\
\hline

Boolean::unsubscribe\_from\_events(\emph{eventType}) & Unsubscribe from event from specific type. 
If event type is not given, unsubscribe from all event types.\\
 & Returns \emph{True} if succeeded; otherwise \emph{False}\\
\hline

\end{tabularx}
\end{table}

%
%
\begin{table}[!ht]
\centering
\caption{NodeProxy API}
\label{api:nodeProxy}
\begin{tabularx}{\textwidth}{@{}XX@{}} \toprule
Function                          & Description     \\ 
\hline

Time::get\_time() & Get time of remote node.\\ 
 & Returns UNIX time of remote node.\\
\hline

Boolean::is\_synchronizing() & Check if remote node is synchronizing with some time server.\\ 
 & Returns \emph{True} is remote node runs time synchronization process; \emph{False} otherwise\\
\hline

String::get\_time\_synchronization\_source() & Get time synchronization source of remote node. Note: we need to check if remote node synchronizes with the same source as application’s local node.\\ 
 & Returns name of synchronization source.\\
\hline

Accuracy::get\_time\_synchronization\_accuracy() & Get time synchronization accuracy.\\ 
 & Returns time synchronization accuracy in milliseconds.\\
\hline

DeviceProxyList::get\_devices() & Get proxy objects for all device modules installed in remote node.\\ 
 & Returns list of Device Module Proxy objects.\\
\hline

DeviceProxy::get\_device(\emph{uuid}) & Get Device Module proxy object by its UUID.\\ 
 & Returns DeviceModuleProxy object.\\
\hline

ProtocolProxyList::get\_protocols() & Get proxy objects for all protocol modules installed in remote node.\\ 
 & Returns list of Protocol Module Proxy objects.\\
\hline

ProtocolProxy::get\_protocol(\emph{uuid}) & Get Protocol Module proxy object by its UUID.\\ 
 & Returns ProtocolModuleProxy object.\\
\hline

ControlApplicationProxyList:: & Get proxy objects for all control\\
get\_control\_applications() & applications installed in remote node.\\ 
 & Returns list of ControlApplicationProxy objects.\\
\hline

ControlApplicationProxy:: & Get Control Application proxy object by\\ get\_control\_application(\emph{uuid}) & its UUID.\\ 
 & Returns ControlApplicationProxy object.\\
\hline

Boolean::send\_event(event) & Send event to remote node in node-broadcast mode, i.e. event is delivered to node and broadcasted to all subscribed Control Applications.\\ 
 & Returns \emph{True} if succeeded; otherwise \emph{False}\\
\hline

Boolean::subscribe\_for\_events(\emph{eventType}, \emph{callback}) & Subscribe for events of given type generated in remote node. If event type is not given, subscribe for all events generated in remote node.  The \emph{callback} function will be called on reception of event.\\ 
 & Returns \emph{True} if succeeded; otherwise \emph{False}\\
\hline

Boolean::unsubscribe\_from\_events(\emph{eventType}) & Unsubscribe from events of given type generated in remote node. If event type is not given unsubscribe from all events generated in remote node.\\ 
 & Returns \emph{True} if succeeded; otherwise \emph{False}\\
\hline

\end{tabularx}
\end{table}

%
%
\begin{table}[!ht]
\centering
\caption{ControlApplicationProxy API}
\label{api:controlAppProxy}
\begin{tabularx}{\textwidth}{@{}XX@{}} \toprule
Function                          & Description     \\ 
\hline

Boolean::is\_running() & Check if remote control application is running.\\ 
 & Returns \emph{True} if remote app is running; otherwise \emph{False}\\
\hline

Boolean::start() & Start remote control application.\\ 
 & Returns \emph{True} if succeeded; otherwise \emph{False}\\
\hline

Boolean::stop() & Stop remote control application.\\ 
 & Returns \emph{True} if succeeded; otherwise \emph{False}\\
\hline

Boolean::send\_event(\emph{event}) & Send event to remote application in unicast mode. Node: event will be delivered only if remote application subscribe for it.\\ 
 & Returns \emph{True} if succeeded; otherwise \emph{False}\\
\hline

Boolean:: & Subscribe for events of given type generated\\
subscribe\_for\_events(\emph{eventType}, \emph{callback}) & in remote control application. If event type is not given, subscribe for all events generated in remote control application. The \emph{callback} function will be called on reception of event.\\ 
 & Returns \emph{True} if succeeded; otherwise \emph{False}\\
\hline

Boolean::unsubscribe\_from\_events(\emph{eventType}) & Unsubscribe from events of given type generated in remote control application. 
If event type is not given unsubscribe from all events generated in remote control application.\\ 
 & Returns \emph{True} if succeeded; otherwise \emph{False}\\
\hline

\end{tabularx}
\end{table}

%
%
\begin{table}[!ht]
\centering
\caption{ModuleProxy API}
\label{api:ModuleProxy}
\begin{tabularx}{\textwidth}{@{}XX@{}} \toprule
Function                          & Description     \\ 
\hline

ModuleProxy::callback(\emph{callbackFunction}) & Execute operation of device in non-blocking mode and register callback function that will be called upon reception of return value from operation.\\ 
 & If \emph{callbackFunction} is not defined operation will be executed in non-blocking mode. \\
 & Returns the same ModuleProxy object -> function chaning.\\
 & \emph{Example:}\\
 & device.callback(\emph{myCallback}).set\_channel(\emph{11}).\\ 
\hline

ModuleProxy::delay(\emph{relativeTime}) & Delay execution of operation by given amount of time. It will result in non-blocking call. Use callback function to register callback.\\
 & Returns the same ModuleProxy object -> function chaning.\\
 & \emph{Example:}\\
 & device.delay(\emph{5s}).set\_channel(\emph{11}).\\ 
\hline

ModuleProxy::exec\_time(\emph{absoluteTime}) & Schedule execution of operation in remote device module. It will result in non-blocking call. Use callback function to register callback. Absolute time is UNIX time. \\
 & Returns the same ModuleProxy object -> function chaning.\\
 & \emph{Example:}\\
 & device.exec\_time(\emph{execTime}).set\_channel(\emph{11}).\\
\hline

Boolean:: & Subscribe for events of given type generated in\\
subscribe\_for\_events(\emph{eventType}, \emph{callback}) & remote device module. If event type is not given, subscribe for all events generated in remote device module. The \emph{callback} function will be called on reception of event.\\
 & Returns \emph{True} if succeeded; otherwise \emph{False}\\
 \hline
 
Boolean:: & Unsubscribe from events of given type generated\\
unsubscribe\_from\_events(\emph{eventType}) & in remote device module. 
If event type is not given unsubscribe from all events generated in remote device module. \\
 & Returns \emph{True} if succeeded; otherwise \emph{False}\\
\hline

\hline
\end{tabularx}
\end{table}

%
%
\begin{table}[!ht]
\centering
\caption{DeviceProxy API}
\label{api:DeviceProxy}
\begin{tabularx}{\textwidth}{@{}XX@{}} \toprule
Function                          & Description     \\ \hline
operation\_1(\emph{arg1,...}) & Python wrapper for remote operation exposed by device.\\ 
operation\_N(\emph{arg1,...}) & Python wrapper for remote operation exposed by device. \\ \hline
\end{tabularx}
\end{table}

%
%
\begin{table}[!ht]
\centering
\caption{ProtocolProxy API}
\label{api:ProtocolProxy}
\begin{tabularx}{\textwidth}{@{}XX@{}} \toprule
Function                          & Description     \\ \hline
operation\_1(\emph{arg1,...}) & Python wrapper for remote operation exposed by device.\\ 
operation\_N(\emph{arg1,...}) & Python wrapper for remote operation exposed by device. \\ \hline
\end{tabularx}
\end{table}

\end{document}